\documentclass[12pt, a4paper]{article}
\usepackage{graphicx}
\usepackage{subeqnarray}
\begin{document}
%
%                         _________________
%                         |               |
%                         |               |
%                         |  VERSIONE  H  |
%                         |               |
%                         |  ...........  |
%                         |               |
%                         |  ...........  |
%                         _________________
%
%
%
\def\astrobj#1{#1}
\newenvironment{lefteqnarray}{\arraycolsep=0pt\begin{eqnarray}}
{\end{eqnarray}\protect\aftergroup\ignorespaces}
\newenvironment{lefteqnarray*}{\arraycolsep=0pt\begin{eqnarray*}}
{\end{eqnarray*}\protect\aftergroup\ignorespaces}
\newenvironment{leftsubeqnarray}{\arraycolsep=0pt\begin{subeqnarray}}
{\end{subeqnarray}\protect\aftergroup\ignorespaces}
\newcommand{\diff}{{\rm\,d}}
\newcommand{\img}{{\rm i}}
\newcommand{\sV}{\mskip 3mu /\mskip-10mu V}
\newcommand{\sP}{\mskip 3mu /\mskip-10mu p}
\newcommand{\sT}{\mskip 3mu /\mskip-08mu T}
\newcommand{\sX}{\mskip 3mu /\mskip-12mu X}
\newcommand{\sY}{\mskip 3mu /\mskip-12mu Y}
\newcommand{\sM}{\mskip 3mu /\mskip-12mu M}
\newcommand{\sA}{\mskip 3mu /\mskip-09mu a}
\newcommand{\appleq}{\stackrel{<}{\sim}}
\newcommand{\appgeq}{\stackrel{>}{\sim}}
\newcommand{\Int}{\mathop{\rm Int}\nolimits}
\newcommand{\Nint}{\mathop{\rm Nint}\nolimits}
\newcommand{\arcsinh}{\mathop{\rm arcsinh}\nolimits}
\newcommand{\range}{{\rm -}}
\newcommand{\sgn}{\mathop{\rm sgn}\nolimits}
\newcommand{\displayfrac}[2]{\frac{\displaystyle #1}{\displaystyle #2}}
\def\astrobj#1{#1}
%\begin{titlepage}
%\setcounter{page}{0}
%\headnote{Astron.~Nachr.~000 (2001) 0, 000--000}
%\makeheadline
%
\title{Tidal interactions and \\ principle of corresponding states: \\
from micro to macro cosmos. \\
A century after van der Waals' Nobel Prize}
\author{{R.~Caimmi}\footnote{
{\it Physics and Astronomy Department, Padua University, Vicolo Osservatorio
3/2, 35122 Padova, Italy.   Affiliated up to September 30th 2014.   Current
status: Studioso Senior.   Current position: in retirement due to age limits.}
\newline email: roberto.caimmi@unipd.it~~~
fax: 39-049-8278212}
%
%, {T.~Valentinuzzi}\footnote{
%{\it Astronomy Department, Padua Univ., Vicolo Osservatorio 2,
%I-35122 Padova, Italy}
%email: tiziano.valentinuzzi@unipd.it~~~
%fax: 39-049-8278212}
%
\phantom{agga}}
%
%\medskip
%\small{Dipartimento di Astronomia}}
%
%\date{Received..................................................
%Accepted..................................................}
\maketitle
\begin{quotation}
\section*{}
\begin{Large}
\begin{center}
%\summary

Abstract

\end{center}
\end{Large}
\begin{small}

\noindent\noindent

The current paper was aimed to honor
the first centennial of Johannes
Diderik van der Waals (VDW) awarding
Nobel Prize in Physics.   VDW
theory of ordinary fluids is reviewed
in the first part of the paper, where
special effort is devoted to the equation
of state and the law of corresponding
states.   In addition, a few mathematical
features involving properties of cubic
equations are discussed, for appreciating
the intrinsic beauty of VDW theory.
A theory of astrophysical fluids is
shortly reviewed in the second part of
the paper, grounding on the tensor virial
theorem for subsystems, and
an equation of state is formulated with
convenient choices of reduced variables.
Additional effort is devoted to selected
density profiles, namely a simple
guidance case and two cases of astrophysical
interest.   Given the analogy between macrogas reduced isoenergetics and
VDW reduced isothermals, a phase transition (gas-stars) is assumed to
take place in astrophysical fluids, similarly to a phase transition
(vapour-liquid) observed in ordinary fluids.   In this framework, the location
of gas clouds, stellar systems, galaxies, cluster of galaxies,
on the plane scanned by reduced
variables, is tentatively assigned.   A brief discussion shows how VDW' two
great discoveries, namely
a gas equation of state where tidal
interactions between molecules are
taken into account and the law of
corresponding states, related to
microcosmos, find a counterpart
with regard to macrocosmos.   In conclusion,
after more than a century since the awarding
of the Nobel Prize in Physics, VDW'
ideas are still valid and
helpful today for a full understanding
of the universe.

\noindent
{\it keywords - gas: ideal, real - gas: equation of state -
galaxies: evolution - dark matter: haloes.}
%END
%\end{titlepage}
\end{small}
\end{quotation}

\section{Introduction} \label{intro}

More than one century ago (1910), the Nobel Prize in Physics was awarded to
Johannes Diderik van der Waals (hereafter quoted as VDW).
In his doctoral thesis (1873) the ideal gas equation of
state was generalized for embracing both the gaseous and
the liquid state, where these two states of aggregation
not only merge into each other in a continuous manner,
but are in fact of the same nature.   With respect to ideal
gases, the volume of the molecules and the intermolecular
tidal forces were taken into account.

VDW equation of state was later
reformulated in terms of reduced (dimensionless) variables (1880), which
allows the description of all substances in terms of a single equation.   In
other words, the state of any substance, defined by the values of
reduced volume, reduced pressure, and reduced temperature, is
independent of the nature of the substance.   This result is
known as the law of corresponding states.

VDW equation of state, in dimensional and reduced form,
served as a guide during experiments which ultimately led
to hydrogen (1898) and helium (1908) liquefaction.   The
Cryogenic Laboratory at Leiden had developed under the
influence of VDW's theories.   For further details on VDW's
biography an interested reader is addressed to specific textbooks (e.g.,
Nobel Lectures 1967).

The current paper was intended to be written in honor of
the first centennial of VDW awarding Nobel Prize in Physics%
\footnote{
It is a revised and improved version of a previous, unpublished paper
available on the arxiv site (arxiv:1210.3688v1).
}.
Ideal and VDW equation of state, both in dimensional and
reduced form, are reviewed, and a number of features are
analysed in detail, in Section \ref{vande}.
Counterparts to ideal and VDW equations of state
for astrophysical fluids, or macrogases, are briefly
summarized and compared with the classical formulation
in Section \ref{macro}.   The discussion and the
conclusion are drawn in Section \ref{disc} and \ref{conc}, respectively.

\section{Equation of state of ordinary fluids}\label{vande}

Let ordinary fluids be conceived as fluids which
can be investigated in laboratory.   The simplest
description is provided by the theory of ideal gas,
where the following restrictive assumptions are made:
(i) particles are identical
spheres; (ii) the number of particles is extremely
large; (iii) the motion of particles is random;
(iv) collisions between particles or with the wall
of the box are perfectly elastic; (v) interactions
between particles or with the wall of the box are
negligible.

Ideal gas equation of state may be written
under the form (e.g., Landau and Lifchitz, 1967,
Chap.\,IV, \S42, hereafter quoted as LL67):
\begin{equation}
\label{eq:gid}
pV=NkT~~;
\end{equation}
where $p$ is pressure, $V$ volume, $T$  temperature, $N$ particle number, and
$k$ Boltzmann constant.

In getting a better description of ordinary fluids,
the above assumption (v) is relaxed and tidal interactions
between particles, due to charge distribution, are taken into consideration.
VDW' generalization
of the equation of state of ideal gases, Eq.\,(\ref{eq:gid}),
reads (van der Waals, 1873):
\begin{equation}
\label{eq:VdW}
\left(p+A\frac{N^2}{V^2}\right)(V-NB)=NkT~~;
\end{equation}
where $A$ and $B$ are constants which depend on particle nature.

More specifically, the presence of an attractive interaction between
particles reduces both force and frequency
of particle-wall collisions: the net effect is a reduction
of pressure, proportional to square numerical
density, expressed as $A(N/V)^2$.   
On the other hand, the whole volume of the box,
$V$, is not accessible to particles, in that they are
represented as identical spheres: the free volume within the
box is $V-NB$, where $NB$ is the covolume.   In particular, $T=0$ implies
$V=NB$, hence the covolume per particle, $B$, may be conceived as the volume
filled by a single sphere in the limit of zero absolute temperature.

Let $2r_0$ be the interparticle distance when the interaction energy is null
(positive values being related to lower distances, implying repulsion, and
negative values to larger distances, implying attraction).  Accordingly, $r_0$
may be conceived as the effective radius of a single sphere, which implies
an effective volume, $V_0=(4\pi/3)r_0^3$.   It can be seen $B$ exceeds $V_0$
by a factor of 4.
For further details, an interested reader is addressed to specific textbooks
(e.g., LL67, Chap.\,\,VII, \S\S72-74).

The features of both ideal and VDW isothermal i.e. constant-temperature
curves, or isothermals, were described in an earlier investigation
(Caimmi 2010, hereafter quoted as C10), to be conceived as the parent paper.
A comparison between VDW and empirical isothermals with equal temperature
shows satisfactory agreement only in presence of a sole phase (gas
or liquid) i.e. for sufficiently low or large
volumes and/or sufficiently high temperatures.
If, on the other hand, two phases
coexist, the pressure of saturated vapour
maintains constant as volume changes, yielding
a horizontal real isothermal.   Real
isothermals are less and less extended for
increasing temperature, until a single point is
attained,
${\sf P}_{\rm c}\equiv(V_{\rm c}, p_{\rm c}, T_{\rm c})$,
which is defined as critical point.   The special VDW isothermal
where the critical point lies, is defined as
critical isothermal.

Parameters appearing in VDW equation of
state, Eq.\,(\ref{eq:VdW}), may be expressed in
terms of coordinates of critical point,
as:
\begin{lefteqnarray}
\label{eq:Vc}
&& V_{\rm c}=3NB~~; \\
\label{eq:Tc}
&& T_{\rm c}=\frac8{27}\frac AB\frac1k~~; \\
\label{eq:pc}
&& p_{\rm c}=\frac1{27}\frac A{B^2}~~; \\
\label{eq:Zc}
&& Z_{\rm c}=\frac{p_{\rm c}V_{\rm c}}{NkT_{\rm c}}=\frac38~~;
\end{lefteqnarray}
where, in general, the compressibility factor:
\begin{lefteqnarray}
\label{eq:cofa}
&& Z=\frac{pV}{NkT}~~;
\end{lefteqnarray}
defines the degree of departure
from the behaviour of ideal gases, for which
$Z=1$, according to Eq.\,(\ref{eq:gid}).
For further details, an interested reader is addressed to specific textbooks
(e.g., Rostagni, 1957, Chap.\,XII, \S20; LL67,
Chap.\,\,VIII, \S85); and to the parent paper (C10).

\begin{figure*}[t]
\begin{center}
\includegraphics[scale=0.8]{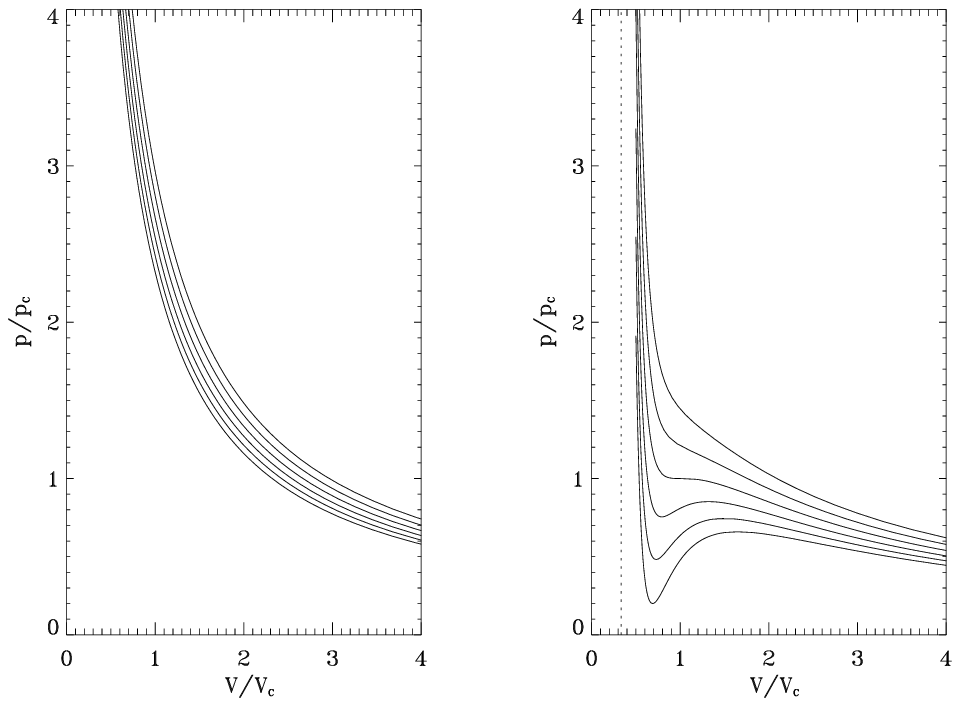}
\caption{Ideal (left panel) and VDW (right panel)
isothermals with equal temperature
(from bottom to top), $T/T_{\rm c}=$
20/23, 20/22, 20/21, 20/20, 20/19, 20/18.
No extremum point exists
above the critical isothermal, $T/T_{\rm c}=1$.
%The dotted line is the common vertical asymptote.
}
\label{f:viso100}
\end{center}
\end{figure*}

\begin{figure*}[t]
\begin{center}
\includegraphics[scale=0.8]{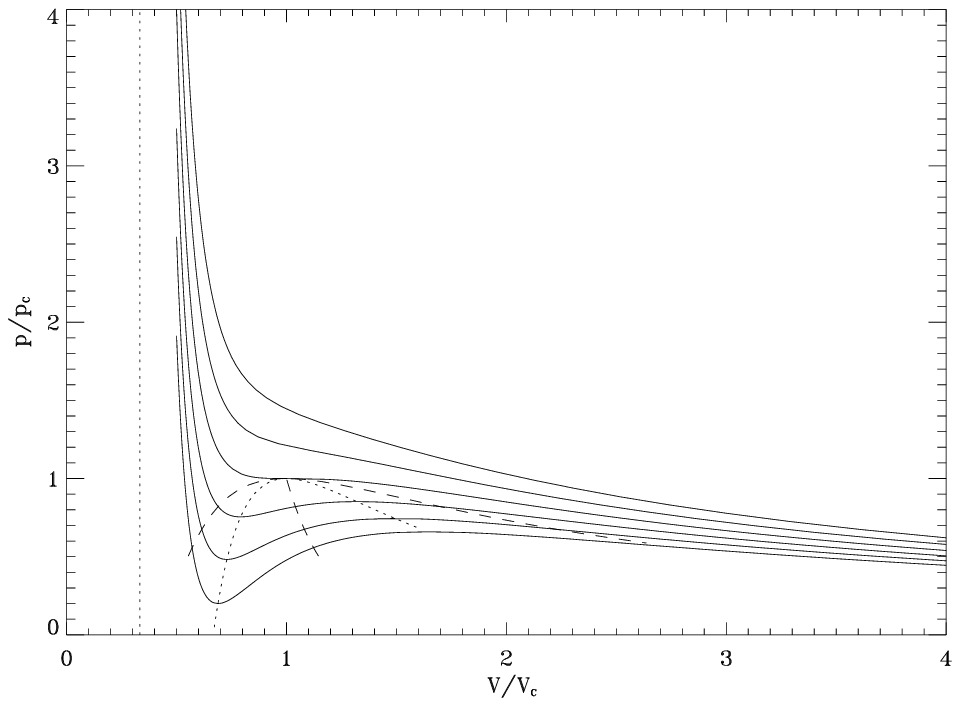}
\caption{Same as in Fig.\,\ref{f:viso100} (right panel), where
the occurrence (within the area bounded by the bell-shaped 
dashed curve) of saturated vapour is considered.
Above the critical isothermal
$(T/T_{\rm c}=1)$ the trend is similar with respect to
ideal gases.   Below the critical isothermal and
on the right of the bell-shaped dashed curve, gas
still behaves as an ideal gas.   Below the critical
isothermal and on the left of the bell-shaped dashed curve,
liquid shows little change in volume as pressure rises.
Within the area bounded by the bell-shaped dashed curve,
liquid phase is in equilibrium with saturated vapour
phase.   A diminished volume implies
smaller saturated vapour fraction and larger liquid fraction
at constant pressure, and vice versa.   VDW equation of state is
no longer valid in this region.   The dashed curve (including
the central branch) is the locus of intersections between VDW
and real isothermals, the latter being related to constant
pressure where liquid and vapour phases coexist.  Ending points are $(1/3,0)$,
left; $(+\infty,0)$, middle; $(+\infty,0)$, right.   The dotted
curve is the locus of VDW isothermal extremum points.   Ending points are
$(1/3,-27)$, left; $(+\infty,0)$, right.}
\label{f:vris100}
\end{center}
\end{figure*}

\begin{figure*}[t]
\begin{center}
\includegraphics[scale=0.8]{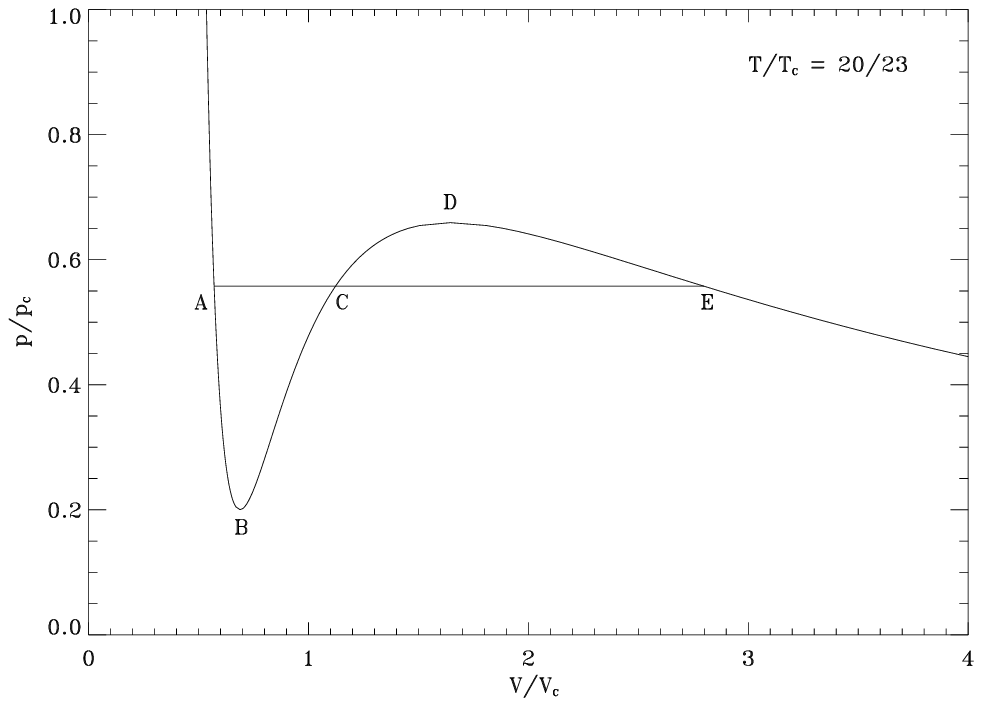}
\caption{VDW and real isothermals with
equal temperature, $T/T_{\rm c}=20/23$.   The two
curves coincide within the range, $V\le V_{\rm A}$
and $V\ge V_{\rm E}$.    VDW isothermal exhibits
two extremum points: minimum, ${\sf B}$, and maximum,
${\sf D}$, while real isothermal is flat within
the range, $V_{\rm A}\le V\le V_{\rm E}$.  Configurations related to
VDW isothermal within the range, $V_{\rm A}\le V\le V_{\rm B}$
(due to tension forces acting on particles, yielding superheated liquid), and
$V_{\rm D}\le V\le V_{\rm E}$ (due to the occurrence of undercooled
vapour), may be obtained under special conditions, while
configurations within the range, $V_{\rm B}\le V\le V_{\rm D}$, are
always unstable.   Volumes, $V_{\rm A}$ and $V_{\rm E}$, correspond
to maximum value in presence of sole liquid phase
and minimum value in presence of sole
vapour phase, respectively.   Regions, {\sf ABC} and
{\sf CDE}, exhibit equal area.}
\label{f:vrar100}
\end{center}
\end{figure*}

Ideal and VDW isothermals with equal
temperature are plotted in Fig.\,\ref{f:viso100}
for different values of $T/T_{\rm c}$ with respect
to the parent paper (C10).
Special regions and loci on the Clapeyron plane,
$({\sf O}Vp)$, are represented in
Fig.\,\ref{f:vris100} for VDW isothermals.
The comparison between real and VDW isothermals with equal
temperature, $T/T_{\rm c}=20/23$, is shown in Fig.\,\ref{f:vrar100}.

For simplifying both notation and calculations,
the (dimensionless) reduced variables are defined
as (e.g., LL67, Chap.\,\,VIII, \S 85):
\begin{equation}
\label{eq:rv}
\sV=\frac V{V_{\rm c}}~~;\qquad\sP=\frac p{p_{\rm c}}~~;
\qquad\sT=\frac T{T_{\rm c}}~~;
\end{equation}
where the coordinates of the critical point are
%${\sf P}_{\rm c}\equiv(V_{\rm c}, p_{\rm c}, T_{\rm c})$,
expressed by Eqs.\,(\ref{eq:Vc}), (\ref{eq:Tc}),
and (\ref{eq:pc}).   Accordingly,
ideal gas equation of state, Eq.\,(\ref{eq:gid}),
and VDW equation of state, Eq.\,(\ref{eq:VdW}),
reduce to:
\begin{lefteqnarray}
\label{eq:ri}
&& \sP\sV=\frac83\sT~~; \\
\label{eq:rW1}
&& \left(\sP+\frac3{\sV^2}\right)\left(\sV-\frac13\right)=\frac83\sT~~;\qquad
\sV>\frac13~~;
\end{lefteqnarray}
respectively.   Additional features related
to VDW isothermals are outlined in
Appendix \ref{a:expo}.   The intersections
between real and VDW isothermals of
equal temperature are analysed and discussed
in Appendix \ref{a:inrv}.   The limit of absolute zero temperature is
considered in Appendix \ref{a:zero}.

The locus of intersections between VDW and real
isothermals of equal temperature is represented in Fig.\,\ref{f:vris100}
as a dashed trifid curve, where the left, the right, and the
middle branch correspond to $\sV_{\rm A}$, $\sV_{\rm E}$, and $\sV_{\rm C}$,
shown in Fig.\,\ref{f:vrar100}, respectively.   The branching point
coincides with the critical point.   The locus of
VDW isothermal extremum points is represented
in Fig.\,\ref{f:vris100} as a dotted bifid curve starting from
the critical point, where the left and the right
branch corresponds to minimum and maximum points, respectively.

A fluid state can be represented in reduced variables
as ($\sV$, $\sP$, $\sT$), where one variable may be
expressed as a function of the remaining two, by use
of ideal gas reduced equation of state, Eq.\,(\ref
{eq:ri}), or VDW reduced equation of state,
Eq.\,(\ref{eq:rW1}).   The formulation in terms of
reduced variables, Eq.\,(\ref{eq:rv}), makes
related equation of state universal i.e. it holds
for any fluid.   Similarly, the Lane-Emden equation
expressed in polytropic (dimensionless) variables
describes the whole class of polytropic gas spheres,
with assigned polytropic index, in hydrostatic
equilibrium (e.g., Chandrasekhar 1939, Chap.\,IV,
\S4).

The coordinates of intersection points between
real and VDW isothermals of equal temperature
and extremum points of VDW isothermals, for
selected temperatures below the critical value, are
listed in Table \ref{t:vispo100} where special effort
has been devoted to the lower neighbourhood of the
critical point.   More specifically, the following dimensionless
parameters have been evaluated vs temperature, $\sT$: lower
volume limit, $\sV_{\rm A}$, where liquid and vapour phase
coexist; extremum point (minimum) volume, $\sV_{\rm B}$;
intermediate volume, $\sV_{\rm C}$, where liquid and vapour phase coexist,
for which pressure equals its counterpart related to corresponding lower and
upper volume limit; extremum point (maximum) volume, $\sV_{\rm D}$;
upper volume limit, $\sV_{\rm E}$, where liquid and
vapour phase coexist; extremum point (minimum) pressure,
$\sP_{\rm B}$; pressure, $\sP_{\rm A}=\sP_{\rm C}=\sP_{\rm E}$,
related to horizontal real isothermal; extremum point
(maximum) pressure, $\sP_{\rm D}$.
\begin{table}
\caption{Values of dimensionless parameters, $\sT$, $\sV_{\rm A}$,
$\sV_{\rm B}$,
$\sV_{\rm C}$, $\sV_{\rm D}$, $\sV_{\rm E}$, $\sP_{\rm B}$, $\sP_{\rm C}$,
$\sP_{\rm D}$, within the range, $0.85\le\sT\le0.99$, using a step,
$\Delta\sT=0.01$.   Additional values are computed near the
critical point, to increase the resolution.
All values equal unity on the critical point.
Index captions: A, C, E - intersections between VDW and real isothermals
of equal temperature; B - extremum point
of minimum; D - extremum point of maximum.   Extremum points
are related to VDW isothermals, while their real
counterparts are flat in presence of both liquid and vapour
phase.   To save aesthetics, 01 on head columns stands for unity and
$9.9\bar9$ on bottom left stands for 9.999.}
\label{t:vispo100}
\begin{center}
\begin{tabular}{|l|l|l|l|l|l|l|l|l|} \hline
$10\sT$ & $10\sV_{\rm A}$ & $10\sV_{\rm B}$ & $01\sV_{\rm C}$ & $01\sV_{\rm D}$ & $01\sV_{\rm E}$ & $10\sP_{\rm B}$ &
$10\sP_{\rm C}$ & $10\sP_{\rm D}$ \\
\hline
 8.50 & 5.5336 & 6.7168 & 1.1453 & 1.7209 & 3.1276 & 0.4963 & 5.0449 & 6.2055 \\
 8.60 & 5.6195 & 6.8003 & 1.1337 & 1.6821 & 2.9545 & 1.2750 & 5.3125 & 6.4005 \\
 8.70 & 5.7116 & 6.8883 & 1.1225 & 1.6436 & 2.7909 & 2.0346 & 5.5887 & 6.6011 \\
 8.80 & 5.8106 & 6.9814 & 1.1116 & 1.6052 & 2.6360 & 2.7752 & 5.8736 & 6.8076 \\
 8.90 & 5.9176 & 7.0804 & 1.1009 & 1.5669 & 2.4889 & 3.4965 & 6.1674 & 7.0205 \\
 9.00 & 6.0340 & 7.1860 & 1.0905 & 1.5285 & 2.3488 & 4.1984 & 6.4700 & 7.2401 \\
 9.10 & 6.1615 & 7.2994 & 1.0804 & 1.4900 & 2.2151 & 4.8807 & 6.7816 & 7.4669 \\
 9.20 & 6.3022 & 7.4221 & 1.0706 & 1.4511 & 2.0869 & 5.5430 & 7.1021 & 7.7014 \\
 9.30 & 6.4593 & 7.5561 & 1.0610 & 1.4117 & 1.9634 & 6.1849 & 7.4318 & 7.9443 \\
 9.40 & 6.6369 & 7.7040 & 1.0516 & 1.3715 & 1.8438 & 6.8058 & 7.7707 & 8.1963 \\
 9.50 & 6.8412 & 7.8697 & 1.0425 & 1.3300 & 1.7271 & 7.4049 & 8.1188 & 8.4584 \\
 9.60 & 7.0819 & 8.0593 & 1.0336 & 1.2867 & 1.6118 & 7.9811 & 8.4762 & 8.7319 \\
 9.70 & 7.3756 & 8.2830 & 1.0249 & 1.2404 & 1.4960 & 8.5328 & 8.8429 & 9.0185 \\
 9.80 & 7.7554 & 8.5611 & 1.0164 & 1.1892 & 1.3761 & 9.0576 & 9.2191 & 9.3209 \\
 9.90 & 8.3091 & 8.9461 & 1.0081 & 1.1278 & 1.2430 & 9.5510 & 9.6048 & 9.6437 \\
 9.95 & 8.7471 & 9.2353 & 1.0040 & 1.0876 & 1.1618 & 9.7830 & 9.8012 & 9.8157 \\
 9.98 & 9.1727 & 9.5049 & 1.0016 & 1.0540 & 1.0972 & 9.9158 & 9.9202 & 9.9240 \\
 9.99 & 9.4018 & 9.6456 & 1.0008 & 1.0377 & 1.0670 & 9.9585 & 9.9600 & 9.9614 \\
 9.9$\bar{9}$ & 9.8035 & 9.8856 & 1.0001 & 1.0117 & 1.0204 & 9.9960 & 9.9960 & 9.9960 \\
%10.0 & 10.000 & 1.0000 & 1.0000 & 1.0000 & 1.0000 & 1.0000 & 1.0000 & 1.0000 \\
\hline
\end{tabular}
\end{center}
\end{table}

Tidal interactions between particles appear in Eq.\,(\ref{eq:VdW}) via
parameters $A$ and $B$.   To gain more insight, the product on the left-hand
side therein can be developed grouping the terms in $A$, $B$, $AB$, and
combining with Eqs.\,(\ref{eq:Vc}), (\ref{eq:pc}), (\ref{eq:cofa}), and
(\ref{eq:rv}).   The result is:
\begin{lefteqnarray}
\label{eq:VdW2}
&& pV\left(1+\frac{9\sV-\sP\sV^2-3}{3\sP\sV^3}\right)=NkT~~; \\
\label{eq:cof2}
&& Z=\frac{pV}{NkT}=\left(1+\frac{9\sV-\sP\sV^2-3}{3\sP\sV^3}\right)^{-1}~~;
\end{lefteqnarray}
where the effect of tidal interactions relates to the second term within
brackets.   More specifically, two additional parameters i.e. critical volume
and critical pressure, appear in the above mentioned term.
Accordingly, VDW equation of state exhibits three variables, $V$, $p$,
$T$, and two parameters, $V_{\rm c}$, $p_{\rm c}$, related to the effects of
tidal interactions between particles.

In terms of reduced variables,
Eq.\,(\ref{eq:VdW2}) via Eq.\,(\ref{eq:Zc}) reads:
\begin{equation}
\label{eq:VrW2}
\sP\sV\left(1+\frac{9\sV-\sP\sV^2-3}{3\sP\sV^3}\right)=\frac83\sT~~;
\end{equation}
where the term within brackets tends to zero as $\sV\to1/3$ i.e. volume
reduces to covolume, which implies null temperature, as expected.

\section{Equation of state of astrophysical fluids}\label{macro}

\subsection{General considerations}\label{gene}

Let macrogases be defined as astrophysical fluids confined by gravitation.
Let two-component system be taken into consideration.
For assigned density profiles, the virial theorem can be
formulated for each subsystem, where the potential
energy is the sum of the self potential energy of
the component under consideration, and the tidal
energy induced by the other one.   Accordingly the virial
theorem for subsystems reads:
\begin{leftsubeqnarray}
\slabel{eq:virsa}
&& 2(E_u)_{\rm kin}+(E_{uv})_{\rm vir}=0~~;\qquad(u,v)=(i,j), (j,i)~~; \\
\slabel{eq:virsb}
&& (E_{uv})_{\rm vir}=(E_{u})_{\rm sel}+(E_{uv})_{\rm tid}~~;
\label{seq:virs}
\end{leftsubeqnarray}
where $i$ and $j$ denote inner and outer subsystem%
\footnote{For sake of simplicity, 
configurations with intersecting boundaries shall
not be considered in the current investigation.},
respectively, $E_{\rm kin}$ is kinetic energy,
$E_{\rm sel}$, $E_{\rm tid}$, and $E_{\rm vir}$
are self, tidal, and virial potential energy,
respectively.   For further details, an interested reader is addressed to
the parent paper (C10).

Astrophysical fluids differ from ordinary fluids
on two main respects, namely (i) the latter are
collisional while the former could be collisionless,
which implies rectilinear and curvilinear
trajectories, respectively, and (ii) macrogases
cannot be bounded by rigid walls, which implies
evaporation.
The assumption of closed systems needs nonzero
pressure on related boundaries, where evaporating
macroparticles are forced to be reflexed as in
rigid walls.

Observables in ordinary fluids (e.g., volume, pressure,
temperature) within cylindric boxes may
be changed acting on a movable circular wall or piston,
by adding or subtracting mechanical energy or
work.  Conceptually, nothing changes if the box
and the piston are thought to be spherical and
the surface of the piston to be variable%
\footnote{A simpler example relates to ordinary
fluids inside conical boxes with a changing-surface
piston moving along the axis of the cone.}.

By analogy, astrophysical fluids must necessarily
be conceived as two-component fluids, where one
subsystem, G, is the macrogas under consideration,
and the other one, P, acts as a piston for changing
the observables.   Energy can be added to or subtracted
from P subsystem (and then to G subsystem via tidal
interaction) by changing the mass, $M_{\rm P}$, and/or
the volume, $V_{\rm P}$, and/or the density profile,
$\rho_{\rm P}$, and determining the virial
equilibrium configuration via the virial theorem for
subsystems.   Let related observables be defined as
macrovolume, $V_{\rm U}$, macropressure, $p_{\rm U}$,
and macrotemperature, $T_{\rm U}$, U = G, P.

Macrovolume is merely the volume filled by
macrogas.   Macropressure has necessarily
dimensions of force per unit area or energy per unit
volume, regardless of (dimensionless) shape factors
and profile factors.
If macroparticles are conceived as mass points,
macrotemperature may be defined as in ordinary fluids:
\begin{equation}
\label{eq:TU}
T_{\rm U}=\frac13\frac{2(E_{\rm U})_{\rm kin}}{N_{\rm U}k}=\frac13\frac
{\overline{m}_{\rm U}(\sigma_{\rm U})^2}k~~;\qquad{\rm U}={\rm G},{\rm P}~~;
\end{equation}
where $N$ is total number of macroparticles,
$\overline{m}=M/N$ mean macroparticle mass, $\sigma$
rms velocity, and $k$ Boltzmann constant.
For typical stellar systems, $\overline{m}=1{\rm m}_
\odot$, $\sigma=144$ km/s, macrotemperature
is $T=10^{62}$ K.  The large value of macrotemperature
with respect to temperature in ordinary fluids arises
from the large value of mean macroparticle mass with
respect to mean particle mass in ordinary fluids.

Strictly speaking, Eq.\,(\ref{eq:TU}) holds for
collisional fluids, where the stress tensor is
isotropic.   For collisionless fluids, the stress
tensor is in general anisotropic.  In any case, it
is diagonal in the reference frame where the
coordinate axes coincide with the principal axes
of inertia.   Accordingly, the macrotemperature
tensor may be defined as:
\begin{lefteqnarray}
\label{eq:TUpp}
&& (T_{\rm U})_{pp}=\frac13\frac{2[(E_{\rm U})_{pp}]_{\rm kin}}{N_{\rm U}k}=
\frac13\frac{\overline{m}_{\rm U}[(\sigma_{\rm U})_{pp}]^2}k~~;\qquad p=1,2,3
~~; \\
\label{eq:sisU}
&& [(\sigma_{\rm U})_{11}]^2+[(\sigma_{\rm U})_{22}]^2+[(\sigma_{\rm U})_{33}]
^2=(\sigma_{\rm U})^2~~;\qquad{\rm U}={\rm G},{\rm P}~~; \\
\label{eq:TsU}
&& (T_{\rm U})_{11}+(T_{\rm U})_{22}+(T_{\rm U})_{33}=T_{\rm U}~~;\qquad
{\rm U}={\rm G},{\rm P}~~;
\end{lefteqnarray}
in the following, attention shall be restricted
to macrotemperature i.e. the trace of macrotemperature tensor.

For assigned subsystems, (U,\,V) = (G,\,P), (P,\,G), the virial
theorem reads:
\begin{equation}
\label{eq:virtU}
-(E_{\rm U})_{\rm sel}-(E_{\rm UV})_{\rm tid}=-(E_{\rm UV})_{\rm vir}=
2(E_{\rm U})_{\rm kin}~~;
\end{equation}
according to Eq.\,(\ref{seq:virs}).
%where $E_{\rm sel}$, $E_{\rm tid}$, are the self and the tidal potential
%energy related to the gravitational potential induced by the subsystem under
%consideration and the other one, respectively; $E_{\rm vir}$ is the virial
%potential energy, defined as the sum of the above two.
For further details, an interested reader is addressed to earlier
investigations
(Limber 1959; Brosche et al. 1983; Caimmi et al. 1984; Caimmi and Secco 1992).

The combination of Eqs.\,(\ref{eq:TU}) and
(\ref{eq:virtU}) yields:
\begin{equation}
\label{eq:virTU}
-\frac13\frac{(E_{\rm U})_{\rm sel}}{V_{\rm U}}V_{\rm U}\left[1+\frac
{(E_{\rm UV})_{\rm tid}}{(E_{\rm U})_{\rm sel}}\right]=N_{\rm U}kT_{\rm U}~~;
\end{equation}
in terms of macrotemperature.
The following definition of macropressure:
\begin{equation}
\label{eq:pU}
p_{\rm U}=-\frac13\frac{(E_{\rm U})_{\rm sel}}{V_{\rm U}}~~;
\end{equation}
translates the virial theorem for subsystems into an equation of state, as:
\begin{equation}
\label{eq:eqsU}
p_{\rm U}V_{\rm U}\left[1+\frac{(E_{\rm UV})_{\rm tid}}{(E_{\rm U})_
{\rm sel}}\right]=N_{\rm U}kT_{\rm U}~~;
\end{equation}
with regard to (U,\,V) macrogases, (U,\,V) = (G,\,P),
(P,\,G).   Accordingly, the ratio:
\begin{equation}
\label{eq:ZU}
Z_{\rm U}=\frac{p_{\rm U}V_{\rm U}}{N_{\rm U}kT_{\rm U}}=\left[1+\frac
{(E_{\rm UV})_{\rm tid}}{(E_{\rm U})_{\rm sel}}\right]^{-1}~~;
\end{equation}
may be conceived as macrogas compressibility factor.   In this view,
Eqs.\,(\ref{eq:eqsU}) and (\ref{eq:ZU}) are macrogas counterparts of
Eqs.\,(\ref{eq:VdW2}) and (\ref{eq:cof2}), respectively, which hold for VDW
gas.

In the limit of an infinitely extended V subsystem, $V_{\rm V}\to+\infty$,
$(E_{\rm UV})_{\rm tid}\to0$, Eq.\,(\ref{eq:eqsU})
has the same formal expression as the ideal gas
equation of state, Eq.\,(\ref{eq:gid}), which, 
in turn, results from the VDW equation of state,
Eq.\,(\ref{eq:VdW}), in absence of tidal action
between molecules.

Strictly speaking, covolume, $N_{\rm U}
B_{\rm U}$, should be taken into consideration,
which should imply $V_{\rm U}>N_{\rm U}B_{\rm U}$
instead of $V_{\rm U}>0$.   In any case, $N_{\rm U}
B_{\rm U}\ll V_{\rm U}$, for astrophysical
fluids, assuming either $B_{\rm G}=V_\odot$,
where $V_\odot$ is the solar volume, or $B_{\rm U}=
(4\pi/3)[(R_{\rm U})_g]^3/N_{\rm U}$, where
$(R_{\rm U})_g=2GM_{\rm U}/c^2$ ($c$ velocity of
light in baryonic matter vacuum) is the gravitational
radius (e.g., Landau and Lifchitz, 1966, Chap.\,IX, \S97) of macrogas under
consideration.  For this reason, covolume shall be neglected in the following.

From this point on, attention shall be restricted to the special case of
homeoidally striated density profiles with similar and similarly placed
boundaries, where coordinate axes coincide with principal axes of inertia.

\subsection{Homeoidally striated density profiles with similar and similarly
placed boundaries}\label{host}

For homeoidally striated density profiles with similar and
similarly placed boundaries, where coordinate axes coincide with
principal axes of inertia, self potential energy and tidal
potential energy take the explicit expression (e.g., C10):
\begin{lefteqnarray}
\label{eq:EUsl}
&& (E_{\rm U})_{\rm sel}=-\frac{G(M_{\rm U})^2}{a_{\rm U}}
\frac{\Xi_{\rm U}(\nu_{\rm U})_{\rm sel}}{[(\nu_{\rm U})_{\rm mas}]^2}\Sigma
~~;\qquad{\rm U}={\rm G},{\rm P}~~; \\
\label{eq:EUVt}
&& (E_{\rm UV})_{\rm tid}=-\frac{G(M_{\rm U})^2}{a_{\rm U}}
\frac{\Xi_{\rm U}(\nu_{\rm UV})_{\rm tid}}{[(\nu_{\rm U})_{\rm mas}]^2}\Sigma
~~;\qquad({\rm U},{\rm V})=({\rm G},{\rm P}), ({\rm P},{\rm G})~~;
\end{lefteqnarray}
where $G$ is constant of gravitation,
$M$ mass, $a=a_1$ truncation radius along the major semiaxis,
$\Xi=\Xi_1$ dimensionless
truncation radius along the major semiaxis, $\nu_{\rm mas}$
and $\nu_{\rm sel}$ are profile
factors i.e. depending only on the density
profile, $\nu_{\rm tid}$ is a
profile factor which, in addition, depends
on the mass ratio, $m=M_{\rm P}/M_{\rm G}$,
and on the homodirection axis ratio, $y=a_{\rm P}/
a_{\rm G}=(a_{\rm P})_r/(a_{\rm G})_r$,
$r=1,2,3$, and $\Sigma$ is a shape factor
i.e. depending only on the boundary, where
$\Sigma=2$ in the special case of spherical
symmetry (e.g., Caimmi and Secco 1992; Caimmi 1993, 1995).

The substitution of Eqs.\,(\ref{eq:EUsl}) and (\ref{eq:EUVt}) into
(\ref{eq:virTU}) after little algebra yields:
\begin{leftsubeqnarray}
\slabel{eq:virFa}
&& \frac13\frac{G(M_{\rm U})^2}{a_{\rm U}}F_{\rm UV}=
N_{\rm U}kT_{\rm U}~~;\qquad({\rm U,V})=
({\rm G},{\rm P}), ({\rm P},{\rm G})~~; \\
\slabel{eq:virFb}
&& F_{\rm UV}=
\frac{\Xi_{\rm U}(\nu_{\rm U})_{\rm sel}}{[(\nu_{\rm U})_{\rm mas}]^2}
\left[1+\frac{(\nu_{\rm UV})_{\rm tid}}{(\nu_{\rm U})_{\rm sel}}\right]
\Sigma~~;
\label{seq:virF}
\end{leftsubeqnarray}
where tidal effects from V subsystem relate to the profile factor,
$(\nu_{\rm UV})_{\rm tid}$.

In addition, Eqs.\,(\ref{eq:pU}), (\ref{eq:eqsU}), and (\ref{eq:ZU}), take the
explicit form:
\begin{lefteqnarray}
\label{eq:pUs}
&& p_{\rm U}=\frac1{4\pi}\frac{G(M_{\rm U})^2}{(a_{\rm U})^4\epsilon_{21}
\epsilon_{31}}\frac{\Xi_{\rm U}(\nu_{\rm U})_{\rm sel}}{[(\nu_{\rm U})_
{\rm mas}]^2}\Sigma~~;\qquad{\rm U}={\rm G},{\rm P}~~; \\
\label{eq:eqsUs}
&& p_{\rm U}V_{\rm U}\left[1+\frac{(\nu_{\rm UV})_{\rm tid}}{(\nu_{\rm U})_
{\rm sel}}\right]=N_{\rm U}kT_{\rm U}~~;\qquad({\rm U},{\rm V})=({\rm G},
{\rm P}), ({\rm P},{\rm G})~~; \\
\label{eq:ZUs}
&& Z_{\rm U}=\frac{p_{\rm U}V_{\rm U}}{N_{\rm U}kT_{\rm U}}=
\left[1+\frac{(\nu_{\rm UV})_{\rm tid}}{(\nu_{\rm U})_{\rm sel}}
\right]^{-1}~;\quad({\rm U},{\rm V})=({\rm G},{\rm P}), ({\rm P},{\rm G})~~;
\end{lefteqnarray}
where $\epsilon_{pq}=a_p/a_q$ are homosurface axis
ratios.   For further details, an interested reader is addressed to the parent
paper (C10), keeping in mind that macropressure therein has a different
definition where neither profile factors nor shape factors are involved.

It is worth remembering Eqs.\,(\ref{eq:eqsUs}) and (\ref{eq:ZUs}) exhibit a
similar formal expression with respect to VDW gases, Eqs.\,(\ref{eq:VdW2}) and
(\ref{eq:cof2}), respectively, where the effects
of tidal interactions relate to the second term within brackets.   More
specifically, two additional parameters i.e. mass ratio and homodirection axis
ratio appear in the above mentioned term.   Accordingly, macrogas equation of
state exhibits three variables, $V_{\rm U}$, $p_{\rm U}$, $T_{\rm U}$, and two
parameters, $m$, $y$, related to the effects of tidal interactions between
subsystems.

For an arbitrary macroisothermal curve,
$T_{\rm U}=$ const, or macroisothermal, the macrogas equation
of state, Eq.\,(\ref{eq:eqsUs}), may be
expressed as:
\begin{equation}
\label{eq:MIU}
\frac{p_{\rm U}V_{\rm U}}{M_{\rm U}}\left[1+\frac{(\nu_{\rm UV})_{\rm tid}}
{(\nu_{\rm U})_{\rm sel}}\right]=\frac{kT_{\rm U}}{\overline{m}_{\rm U}}~~;
\qquad({\rm U},{\rm V})=({\rm G},{\rm P}),({\rm P},{\rm G})~~;
\end{equation}
where $0<V_{\rm U}<+\infty$ and
$(\nu_{\rm UV})_{\rm tid}/(\nu_{\rm U})_{\rm sel}\ge0$
in the case under consideration of
homeoidally striated density profiles
with similar and similarly placed
boundaries, as shown below.   For homogeneous spherical
configurations, $\Xi=1$, $\nu_{\rm mas}=
\Xi^3=1$, $\nu_{\rm sel}=3\Xi^5/10=3/10$,
$\epsilon_{pq}=1$, $\Sigma=2$, and
Eq.\,(\ref{eq:pUs}) reduces to:
\begin{lefteqnarray}
\label{eq:pUhsp}
&& (p_{\rm U})_{\rm hsp}=\frac1{4\pi}\frac35\frac{G(M_{\rm U})^2}
{(a_{\rm U})^4}~~;
\end{lefteqnarray}
where the index, hsp, means homogeneous sphere.
Related self potential energy, via
Eq.\,(\ref{eq:pU}), is:
\begin{lefteqnarray}
\label{eq:EUhsp}
&& [(E_{\rm U})_{\rm sel}]_{\rm hsp}=-\frac35\frac{G(M_{\rm U})^2}
{a_{\rm U}}~~;
\end{lefteqnarray}
which may be considered as a limiting
value for the following reasons.

Gravitational potential energy of a matter
distribution (in absolute value) may be
conceived as a binding energy which,
{\it ipso facto}, is increasing
for increasing concentration at fixed
total mass and major semiaxis.   Accordingly,
both inhomogeneous and aspherical configurations
increase the binding energy provided density
profiles with positive slope are excluded as
unphysical.   In this view, homogeneous
spherical configurations are related to
a minimum binding energy, $-[(E_{\rm U})_
{\rm sel}]_{\rm hsp}\le-(E_{\rm U})_{\rm sel}$.

On the other hand, the tidal action exerted from the external homeoid of the
outer subsystem on the inner one is null owing to Newton's theorem, which
implies
$(E_{ij})_{\rm tid}/(E_i)_{\rm sel}=(\nu_{ij})_{\rm tid}/(\nu_i)_{\rm sel}>0$.
In addition, the tidal action exerted from the inner subsystem on the outer
one implies
$(E_{ji})_{\rm tid}/(E_j)_{\rm sel}=(\nu_{ji})_{\rm tid}/(\nu_j)_{\rm sel}>0$.
This is why the presence of a second subsystem enhances binding energy
with respect to a single subsystem.   In conclusion,
$(E_{\rm UV})_{\rm tid}/(E_{\rm U})_{\rm sel}=(\nu_{\rm UV})_{\rm tid}/
(\nu_{\rm U})_{\rm sel}>0$, (U,V) = (G,P), (P,G).

According to the above considerations, the left-hand side of
Eq.\,(\ref{eq:eqsUs}) via (\ref{eq:virTU})-(\ref{eq:eqsU}) satisfies the
following condition:
\begin{lefteqnarray}
\label{eq:pVsph}
&& p_{\rm U}V_{\rm U}\left[1+\frac{(\nu_{\rm UV})_{\rm tid}}{(\nu_{\rm U})_
{\rm sel}}\right]=-\frac13[(E_{\rm U})_{\rm sel}+(E_{\rm UV})_{\rm tid}]
\nonumber \\
&& =%\phantom{p_{\rm U}V_{\rm U}\left[1+\frac{(\nu_{\rm UV})_{\rm tid}}{(\nu_{\rm U})_{\rm sel}}\right]}=
\frac13\frac{G(M_{\rm U})^2}{a_{\rm U}}\frac{\Xi_{\rm U}(\nu_{\rm U})_
{\rm sel}}{[(\nu_{\rm U})_{\rm mas}]^2}\left[1+\frac{(\nu_{\rm UV})_{\rm tid}}
{(\nu_{\rm U})_{\rm sel}}\right]\Sigma>\frac13\frac35\frac{G(M_{\rm U})^2}
{a_{\rm U}}~~;
\end{lefteqnarray}
or:
\begin{equation}
\label{eq:sph}
\frac{\Xi_{\rm U}(\nu_{\rm U})_{\rm sel}}{[(\nu_{\rm U})_{\rm mas}]^2}
\left[1+\frac{(\nu_{\rm UV})_{\rm tid}}{(\nu_{\rm U})_{\rm sel}}\right]
\Sigma>\frac35~~;
\end{equation}
which depends on the density profiles (including shape) and, in addition, on
the homodirection
axis ratio, $y$, and the mass ratio, $m$, via $(\nu_{\rm UV})_{\rm tid}$.

Let $(V_{\rm U},p_{\rm U})$ be an assigned state on
a selected macroisothermal, $T_{\rm U}=$ const.
An increasing macrovolume, $V_{\rm U}$, or major semiaxis,
$a_{\rm U}$, at constant mass, $M_{\rm U}$, implies
a decreasing left-hand side of Eq.\,(\ref{eq:eqsUs}),
via Eq.\,(\ref{eq:pUs}), which may be compensated by
increasing concentration and/or asphericity
via Eq.\,(\ref{eq:sph}), with due account taken of a
changing tidal interaction due to the factor, $[1+(\nu_
{\rm UV})_{\rm tid}/(\nu_{\rm U})_{\rm sel}]$, appearing
in Eq.\,(\ref{eq:eqsUs}), until an infinite macrovolume
is attained.   In particular, $[\Xi\nu_{\rm sel}/
(\nu_{\rm mas})^2]\Sigma=3/(5-n)$ for polytropic spheres
of polytropic index equal to $n$ $(0\le n\le5)$,
$\Sigma=\pi$ for flattened oblate spheroids $(\epsilon
_{21}=1$, $\epsilon_{31}\to0)$, and $\Sigma\to+\infty$
for elongated prolate spheroids $(\epsilon_{21}=\epsilon_{31}\to0)$.

A decreasing macrovolume, $V_{\rm U}$, or major semiaxis,
$a_{\rm U}$, at constant mass, $M_{\rm U}$, implies
an increasing left-hand side of Eq.\,(\ref{eq:eqsUs}),
via Eq.\,(\ref{eq:pUs}), which may be compensated by
decreasing concentration and/or asphericity
via Eq.\,(\ref{eq:sph}), with due account taken of a
changing tidal interaction via the factor, $[1+(\nu_
{\rm UV})_{\rm tid}/(\nu_{\rm U})_{\rm sel}]$, appearing
in Eq.\,(\ref{eq:eqsUs}), until a minimum macrovolume
is attained, which makes the ending point of the
macroisothermal.   In fact, lower macrovolumes
would imply density profiles with positive slope,
unless the assumption of mass conservation is
released.   By increasing mass, a macroisothermal
can be continued until null macrovolume (or
volume equal to covolume) is asymptotically attained.

Accordingly, astrophysical fluids lying on
macroisothermals must be conceived as
open systems i.e. variable macroparticle number,
contrary to ordinary fluids, which can be conceived
as closed systems i.e. constant particle number.
Then a more germane formulation of macrogas equation
of state, Eq.\,(\ref{eq:MIU}), reads:
\begin{equation}
\label{eq:RIU}
\frac{p_{\rm U}}{\overline{\rho}_{\rm U}}\left[1+\frac{(\nu_{\rm UV})_
{\rm tid}}{(\nu_{\rm U})_{\rm sel}}\right]=\frac{kT_{\rm U}}{\overline{m}_
{\rm U}}~~;\qquad({\rm U,V})=({\rm G,P}), ({\rm P,G})~~;
\end{equation}
where mean density, $\overline{\rho}=M/V$,
appears instead of total mass, $M$.

Dividing both sides of Eq.\,(\ref{eq:eqsUs})
where (U,\,V) = (P,\,G), by their counterparts where
(U,\,V) = (G,\,P), yields:
\begin{equation}
\label{eq:eqPG}
Y_{\rm p}Y_{\rm V}\frac{1+(\nu_{\rm PG})_{\rm tid}/(\nu_{\rm P})_{\rm sel}}
{1+(\nu_{\rm GP})_{\rm tid}/(\nu_{\rm G})_{\rm sel}}
=\frac{N_{\rm P}}{N_{\rm G}}Y_{\rm T}~~;
\end{equation}
in terms of the dimensionless variables:
\begin{equation}                                                
\label{eq:YpVT}
Y_{\rm p}=\frac{p_{\rm P}}{p_{\rm G}}~~;\qquad
Y_{\rm V}=\frac{V_{\rm P}}{V_{\rm G}}~~;\qquad
Y_{\rm T}=\frac{T_{\rm P}}{T_{\rm G}}~~;
\end{equation}
or, using Eqs.\,(\ref{eq:TU}) and
(\ref{eq:pUs}):
\begin{eqnarray}
\label{eq:Ymsp}
&& Y_{\rm p}=\frac{m^2}{y^4}\frac{\Xi_{\rm P}}{\Xi_{\rm G}}\frac{(\nu_{\rm P})_
{\rm sel}}{(\nu_{\rm G})_{\rm sel}}\left[\frac{(\nu_{\rm G})_{\rm mas}}
{(\nu_{\rm P})_{\rm mas}}\right]^2~~;\qquad Y_{\rm V}=y^3~~;\qquad
Y_{\rm T}=\frac{N_{\rm G}}{N_{\rm P}}\phi~~;\qquad \\
\label{eq:msp}
&& m=\frac{M_{\rm P}}{M_{\rm G}}~~;\qquad y=\frac{a_{\rm P}}{a_{\rm G}}~~;
\qquad\phi=\frac{(E_{\rm P})_{\rm kin}}{(E_{\rm G})_{\rm kin}}=
\frac{(E_{\rm PG})_{\rm vir}}{(E_{\rm GP})_{\rm vir}}~~;
\end{eqnarray}
where the mass ratio, $m$, has not to be confused
with the mean macroparticle mass, $\overline{m}$.
The combination of Eqs.\,(\ref{eq:TU}) and
(\ref{eq:YpVT})-(\ref{eq:msp}) yields:
\begin{equation}
\label{eq:TPG}
T_{\rm P}=\frac{\overline{m}_{\rm P}}{\overline{m}_{\rm G}}\frac\phi m
T_{\rm G}~~;
\end{equation}
where $\overline{m}_{\rm P}=\overline{m}_{\rm G}$
without loss of generality, provided P macrogas is
considered only for the tidal potential induced on
G macrogas.

Owing to Eqs.\,(\ref{eq:Ymsp}) and (\ref{eq:msp}),
Eq.\,(\ref{eq:eqPG}) by use of (\ref{eq:virFb})
takes the explicit form:
\begin{equation}
\label{eq:eqse}
\frac{m^2}y\frac{F_{\rm PG}}{F_{\rm GP}}
%\frac{\Xi_{\rm P}}{\Xi_{\rm G}}\frac{(\nu_{\rm P})_
%{\rm sel}}{(\nu_{\rm G})_{\rm sel}}\left[\frac{(\nu_{\rm G})_{\rm mas}}
%{(\nu_{\rm P})_{\rm mas}}\right]^2\frac{1+(\nu_{\rm PG})_{\rm tid}/
%(\nu_{\rm P})_{\rm sel}}{1+(\nu_{\rm GP})_{\rm tid}/(\nu_{\rm G})_{\rm sel}}
=\phi~~;
\end{equation}
in terms of the dimensionless variables (C10):
\begin{equation}                                                
\label{eq:XpVT}
X_{\rm p}=m^2~~;\qquad X_{\rm V}=\frac1y~~;\qquad X_{\rm T}=\phi~~;
\end{equation}
where the dimensionless factors, $F$,
depend on the density profiles and,
in addition, on the variables, $m$,
$y$, via the profile factors, $\nu_
{\rm tid}$.

In the case under consideration of homeoidally striated, similar and similarly
placed ellispoids, the dependence on the shape implies the same factor,
$\Sigma$, for both  $F_{\rm PG}$ and $F_{\rm GP}$, which disappears in the
ratio, $F_{\rm PG}/F_{\rm GP}$.   Accordingly, Eq.\,(\ref{eq:eqse}) may be
conceived as shape independent.

The additional restriction of coinciding
scaled density profiles, $f_{\rm U}=\rho_{\rm U}/\rho_{\rm U}^\dagger$,
$\rho_{\rm U}^\dagger=\rho_{\rm U}(r_{\rm U}^\dagger)$, $r_{\rm U}^\dagger$
selected scaling radius along a fixed direction, U = G,P; coinciding scaled
truncation radii, $\Xi_{\rm U}=R_{\rm U}/r_{\rm U}^\dagger$, $R_{\rm U}$
truncation radius along the same direction, U = G,P; and coinciding
homodirection axes, $y=1$, makes Eq.\,(\ref{eq:eqse}) reduce to:
\begin{equation}                                                
\label{eq:mfy1}
m=\phi~~;\qquad y=1~~;
\end{equation}
for a formal demonstration, an interested reader is addressed to the parent
paper (C10).

Due to appearence of parameters from both G and P
macrogases, Eq.\,(\ref{eq:eqPG}) or
equivalently Eq.\,(\ref{eq:eqse}) may be
conceived as a macrogas reduced
equation of state, depending only on
dimensionless parameters.
More specifically, macrogas reduced equation of state, Eq.\,(\ref{eq:eqse}),
depends on three
dimensionless variables, namely mass ratio, $m$, homodirection axis ratio,
$y$, and kinetic energy ratio, $\phi$, which is the counterpart of VDW reduced
equation, Eq.\,(\ref{eq:VrW2}), for VDW gases.

For assigned macrogases, the locus,
$\phi=$ const, on the dimensionless
plane, $({\sf O}y^{-1}m^2)$, may be conceived as a fractional
isoenergetic curve, or fractional isoenergetic, in the sense that
kinetic energy ratio, $\phi$, instead of
macrotemperature ratio, $Y_{\rm T}$,
maintains constant.   It can be seen the latter alternative implies a
subdomain where configurations have no physical meaning.   For further
details, an interested reader is addressed to Appendix \ref{a:remi}. 

A generic point, $(X_{\rm V},X_{\rm p})$, on a fractional isoenergetic,
$X_{\rm T}$,
is connected to an infinity of configurations where homodirection axis ratio,
$y$, mass ratio, $m$, kinetic energy ratio, $\phi$, remain unchanged.
Accordingly, macrogas equations of state, Eqs.\,(\ref{seq:virF}), are
``entangled'': for fixed
$a_{\rm U}$, $M_{\rm U}$, $(E_{\rm U})_{\rm kin}=(3/2)N_{\rm U}kT_{\rm U}$,
related counterparts
$a_{\rm V}$, $M_{\rm V}$, $(E_{\rm V})_{\rm kin}=(3/2)N_{\rm V}kT_{\rm V}$,
are inferred via $y$, $m$, $\phi$, by use of Eq.\,(\ref{eq:msp}).   Keeping in
mind P subsystem is conceived as a piston for changing the state of the other
one, attention shall mainly be restricted to G subsystem.

With regard to a generic curve on $({\sf O}X_{\rm V}X_{\rm p})$ plane, and a
generic subdomain, $X_{\rm V_1}\le X_{\rm V}\le X_{\rm V_2}$, related area
below the curve reads:
\begin{lefteqnarray}
\label{eq:area}  
&& {\cal S}_{12}=\int_{X_{\rm V_1}}^{X_{\rm V_2}}X_{\rm p}\diff X_{\rm V}=
\int_{y_1^{-1}}^{y_2^{-1}}m^2\diff\frac1y=\int_{y_2}^{y_1}\frac{m^2}{y^2}\diff
y~~;
\end{lefteqnarray}
where, in particular, mass, $M_{\rm G}$, and radius, $a_{\rm G}$, can be left
unchanged while $M_{\rm P}$ and $a_{\rm P}$ vary via Eqs.\,(\ref{eq:msp}) and
(\ref{eq:XpVT}).
Accordingly, Eq.\,(\ref{eq:area}) translates into:
\begin{lefteqnarray}
\label{eq:arec}  
&& {\cal S}_{12}\frac{GM_{\rm G}^2}{a_{\rm G}}=\int_{a_{\rm P_2}}^
{a_{\rm P_1}}\frac{GM_{\rm P}^2}{a_{\rm P}^2}\diff a_{\rm P}~~;
\end{lefteqnarray}
which may be conceived as the work performed on G macrogas via P macrogas to
leave $M_{\rm G}$ and $a_{\rm G}$ unchanged passing from $(M_{\rm P_1},
a_{\rm P_1})$ to $(M_{\rm P_2}, a_{\rm P_2})$.

A similar result holds for a generic curve on the
$({\sf O}X_{\rm V}^{-1}X_{\rm p}^{-1})$ plane, where the counterpart of
Eq.\,(\ref{eq:arec}) exhibiting indexeses, G and P, reversed,
may be conceived as the work performed on P macrogas via G macrogas to
leave $M_{\rm P}$ and $a_{\rm P}$ unchanged passing from $(M_{\rm G_1},
a_{\rm G_1})$ to $(M_{\rm G_2}, a_{\rm G_2})$.

Macrogas equation of state, Eq.\,(\ref{eq:eqsUs}) via Eqs.\,(\ref{eq:TU}) and
(\ref{seq:virF})
exhibits three variables, $a_{\rm G}$, $M_{\rm G}$, $(E_{\rm G})_{\rm kin}$,
and two parameters, $m$, $y$, which are fixed by the point selected on the
fractional isoenergetic, where a third parameter, $\phi$, remains unchanged.
If, in addition, attention is restricted to isoenergetics,
$(E_{\rm G})_{\rm kin}=(3/2)N_{\rm G}kT_{\rm G}={\rm const}$, where
mass conservation, $M_{\rm G}={\rm const}$, $N_{\rm G}={\rm const}$, takes
place, then $T_{\rm G}={\rm const}$, and the remaining variable, $a_{\rm G}$,
can be determined via Eq.\,(\ref{seq:virF}).

It is worth emphasizing fractional isoenergetics, $X_{\rm T}={\rm const}$, and
fractional
macroisothermals, $Y_{\rm T}=(N_{\rm G}/N_{\rm P})\phi=(\overline m_{\rm P}/
\overline m_{\rm G})(\phi/m)={\rm const}$ via Eq.\,(\ref{eq:TPG}), are
equivalent provided mass conservation takes place in both macrogases.   The
same holds for isoenergetics,
$(E_{\rm U})_{\rm kin}=(3/2)N_{\rm U}kT_{\rm U}={\rm const}$ via
Eq.\,(\ref{eq:TU}), and macroisothermals, $T_{\rm U}={\rm const}$.

For assigned shape, density profiles, and truncation radii,
isoenergetics or macroisothermals can be determined along the following steps.
\begin{description}
\item [(i)] Start from a generic configuration among $(m, y)$.
\item [(ii)] For assigned $\phi=X_{\rm T}$, via Eq.\,(\ref{eq:eqse}) plot
related fractional isoenergetic on $({\sf O}X_{\rm V}X_{\rm p})$ plane.
\item [(iii)] For assigned mass, $M_{\rm G}$, and kinetic energy,
$(E_{\rm G})_{\rm kin}=N_{\rm G}kT_{\rm G}$, via Eq.\,(\ref{seq:virF})
determine major semiaxis, $a_{\rm G}$ and via Eq.\,(\ref{eq:msp}) determine
$M_{\rm P}$, $a_{\rm P}$, $(E_{\rm P})_{\rm kin}=N_{\rm P}kT_{\rm P}$.
\item [(iv)] Return to (iii) and change $m$, $y$, along the fractional
isoenergetic leaving $M_{\rm G}$, $N_{\rm G}kT_{\rm G}$, unchanged.
\item [(v)] Using Eq.\,(\ref{eq:pUs}), determine $p_{\rm G}$, $V_{\rm G}$;
$p_{\rm P}$, $V_{\rm P}$; and plot related isoenergetics on
$({\sf O}V_{\rm G}p_{\rm G})$; $({\sf O}V_{\rm P}p_{\rm P})$; plane,
respectively.
\item [(vi)] Return to (ii) and repeat the procedure until a family of
fractional
isoenergetics is plotted on $({\sf O}X_{\rm V}X_{\rm p})$ plane, and a
family of isoenergetics is plotted on
$({\sf O}V_{\rm G}p_{\rm G})$ and $({\sf O}V_{\rm P}p_{\rm P})$ plane for
different $N_{\rm G}kT_{\rm G}$ and $N_{\rm P}kT_{\rm P}$, respectively.
\end{description}
Accordingly, a selected isoenergetic on the
plane, $({\sf O}V_{\rm G}p_{\rm G})$,
Eq.\,(\ref{eq:eqsUs}), corresponds to a fractional
isoenergetic on the dimensionless
plane, $({\sf O}X_{\rm V}X_{\rm p})$,
Eq.\,(\ref{eq:eqse}), and vice versa,
conformly to Eqs.\,(\ref{eq:YpVT})-(\ref{eq:TPG}).
%Steps (iv) and (v) above would not be possible all over the whole domain,
%$y>0$, if $X_{\rm T}=\phi/m=$ const is assumed instead of $\phi=$ const,
%which is the reason for the current choice.

The procedure outlined above defines a
special way of making macroiso\-thermal
changes of state, $(V_{\rm G}, p_{\rm G},
T_{\rm G})\to(V_{\rm G}+\Delta V_{\rm G},
p_{\rm G}+\Delta p_{\rm G},T_{\rm G})$,
and in this sense it is not restrictive.
With regard to ordinary fluids, it is
equivalent to vary mechanical and
thermal energy by assigned amounts
making isothermal changes of state,
$(V, p, T)\to(V+\Delta V, p+\Delta p, T)$.

The definition of dimensionless variables, $Y_{\rm V}, Y_{\rm p}, Y_{\rm T}$,
expressed by Eq.\,(\ref{eq:YpVT}), aims to a closer analogy between ordinary
and astrophysical fluids with respect
to the parent paper (C10) where attention was restricted to $X_{\rm V},
X_{\rm p}, X_{\rm T}$, expressed by Eq.\,(\ref{eq:XpVT}),
and fractional isoenergetics were conceived as macroiso\-thermals.

Fractional isoenergetics were derived in the parent paper (C10) for a wide
amount
of cases, with regard to different kind of macrogases, namely UU, HH, and
HN/NH.   More specifically, UU is a simple guidance case where no critical
point occurs; conversely, HH and HN/NH are cases of astrophysical interest
where critical points take place.   Accordingly, isoenergetics can be inferred
from the above mentioned fractional isoenergetics.

Critical fractional isoenergetics, hosting a critical point i.e. a horizontal
inflexion point, take place in macrogases for sufficiently steep density
profiles: fractional isoenergetics on one side exhibit two extremum points
(maximum and minimum), while fractional isoenergetics on the other side
exhibit
no extremum point, similarly to reduced VDW isothermals.   On the contrary,
critical fractional isoenergetics are absent for sufficiently mild density
profiles, where fractional isoenergetics show a nonmonotonic trend.
For further details, an interested reader is addressed to the parent
paper (C10) and an earlier investigation (Caimmi and Valentinuzzi 2008).

The existence of a phase transition moving along
a selected fractional isoenergetic in presence of
extremum points (model fractional isoenergetic),
where the path is a horizontal line
(transition fractional isoenergetic),
must necessarily be assumed as a working
hypothesis, due to the analogy between VDW isothermals
and model fractional isoenergetics.   Unlike VDW equation of state,
Eq.\,(\ref{eq:VdW}), macrogas equation of state,
Eq.\,(\ref{eq:MIU}), is not analytically integrable, which implies
the procedure, used for determining a selected
fractional isoenergetic, must be numerically performed.

The main steps are the following.
\begin{description}
\item [(i)] Calculate intersections,
$V_{\rm GA}$, $V_{\rm GC}$, $V_{\rm GE}$;
$V_{\rm GA}<V_{\rm GC}<V_{\rm GE}$; between the
generic horizontal line on the plane,
$({\sf O}V_{\rm G}p_{\rm G})$, $p_{\rm G}=$
const, and the curve related to theoretical macrogas
equation of state, within the range, $p_{\rm GB}<p_{\rm G}<
p_{\rm GD}$, where B and D denote extremum points
of minimum and maximum, respectively.
\item [(ii)] Calculate area of regions, ${\sf ABC}$ and ${\sf CDE}$.
\item [(iii)]
Find the special value, $p_{\rm G}=p_{\rm GC}$, which makes regions of equal
area.
\item [(iv)] Trace the transition fractional isoenergetic
as a horizontal line connecting points,
$(V_{\rm GA},p_{\rm GA})$, $(V_{\rm GC},p_{\rm GC})$,
$(V_{\rm GE},p_{\rm GE})$, $p_{\rm GA}=p_{\rm GC}=
p_{\rm GE}$.
\end{description}

For further details, an interested reader is addressed to the
parent paper (C10).   In order to preserve the
analogy with ideal and VDW gases, the tidal potential energy has to be
negligible and comparable, respectively, with regard to self potential energy,
in the formulation of the virial theorem for subsystems and related equation
of state concerning macrogas of interest.

For assigned density profiles allowing critical
macroisothermal, Eqs.\,(\ref{eq:eqsUs}), (U, V)
= (G, P), (\ref{eq:eqPG}), and  (\ref{eq:eqse}), may be translated
into reduced variables, as:
\begin{lefteqnarray}
\label{eq:eqrG}  
&& \sP_{\rm G}\sV_{\rm G}\frac{1+(\nu_{\rm GP})_{\rm tid}/(\nu_{\rm G})_
{\rm sel}}{1+[(\nu_{\rm GP})_{\rm tid}]_{\rm c}/(\nu_{\rm G})_{\rm sel}}=
\frac{M_{\rm G}}{(M_{\rm G})_{\rm c}}\sT_{\rm G}~~; \\
\label{eq:rG}  
&& \sP_{\rm G}=\frac{p_{\rm G}}{(p_{\rm G})_{\rm c}}~~;\qquad
\sV_{\rm G}=\frac{V_{\rm G}}{(V_{\rm G})_{\rm c}}~~;\qquad
\sT_{\rm G}=\frac{T_{\rm G}}{(T_{\rm G})_{\rm c}}~~; \\
\label{eq:eaGP}
&& \sY_{\rm p}\sY_{\rm V}F_{\rm c}=\frac m{m_{\rm c}}\sY_{\rm T}~~; \\
\label{eq:aGP}
&& \sY_{\rm p}=\frac{Y_{\rm p}}{Y_{\rm p_{\rm c}}}~~;\qquad
   \sY_{\rm V}=\frac{Y_{\rm V}}{Y_{\rm V_{\rm c}}}~~;\qquad
   \sY_{\rm T}=\frac{Y_{\rm T}}{Y_{\rm T_{\rm c}}}~~; \\
\label{eq:FX}
&& F_{\rm c}=\frac{1+(\nu_{\rm PG})_{\rm tid}/(\nu_{\rm P})_ {\rm sel}}
     {1+[(\nu_{\rm PG})_{\rm tid}]_{\rm c}/(\nu_{\rm P})_{\rm sel}}
\frac{1+[(\nu_{\rm GP})_{\rm tid}]_{\rm c}/(\nu_{\rm G})_{\rm sel}}
     {1+(\nu_{\rm GP})_{\rm tid}/(\nu_{\rm G})_ {\rm sel}}~~; \\
\label{eq:FX}
&& \sX_{\rm p}\sX_{\rm V}F_{\rm c}=\sX_{\rm T}~~; \\
&& \sX_{\rm p}=\frac{X_{\rm p}}{\sX_{\rm p_c}}~~;\qquad
   \sX_{\rm V}=\frac{X_{\rm V}}{\sX_{\rm V_c}}~~;\qquad
   \sX_{\rm T}=\frac{X_{\rm T}}{\sX_{\rm T_c}}~~;\qquad 
\end{lefteqnarray}
where the index, c, denotes critical point, and dimensionless variables,
$X$, are defined by Eq.\,(\ref{eq:XpVT}).

In the limit of an infinitely extended P subsystem, $y\to+\infty$,
$(\nu_{\rm PG})_{\rm tid}\to0$, $(\nu_{\rm GP})_{\rm tid}\to0$,
Eqs.\,(\ref{eq:eqrG}) and (\ref{eq:eaGP}) reduce to:
\begin{lefteqnarray}
\label{eq:eirG}  
&& \sP_{\rm G}\sV_{\rm G}
%\frac{(p_{\rm G})_{\rm c}(V_{\rm G})_{\rm c}}{(T_{\rm G})_{\rm c}}
=\left\{1+\frac{[(\nu_{\rm GP})_{\rm tid}]_{\rm c}}{(\nu_{\rm G})_{\rm sel}}
\right\}\frac{M_{\rm G}}{(M_{\rm G})_{\rm c}}\sT_{\rm G}~~; \\
\label{eq:iaGP}
&& \sY_{\rm p}\sY_{\rm V}
%\frac{X_{\rm p_{\rm c}}X_{\rm V_{\rm c}}}{X_{\rm T_{\rm c}}}
=\frac{1+[(\nu_{\rm PG})_{\rm tid}]_{\rm c}/(\nu_{\rm P})_{\rm sel}}
{1+[(\nu_{\rm GP})_{\rm tid}]_{\rm c}/(\nu_{\rm G})_{\rm sel}}
\frac m{m_{\rm c}}\sY_{\rm T}~~;
\end{lefteqnarray}
which can be related to ordinary gases filling a fixed volume. For further
details, an interested reader is addressed to the parent paper (C10).
%The following macrogases shall be dealt with: UU, HH, and HN/NH.   More
%specifically, UU is a simple guidance case where no critical point occurs;
%conversely, HH and HN/NH are cases of astrophysical interest where critical
%points take place.

The above result, expressed by Eq.\,(\ref{eq:eqrG}), may be written in
parametric form using Eqs.\,(\ref{eq:pUs}) and (\ref{eq:eqsUs}), as:
\begin{lefteqnarray}
\label{eq:pVTc}  
&& \sP_{\rm G}=\frac{\sM_{\rm G}^2}{\sA_{\rm G}^4}~~;\qquad\sV_{\rm G}=
\sA_{\rm G}^3~~;\qquad\sT_{\rm G}=\frac{\sM_{\rm G}}{\sA_{\rm G}}\chi_{\rm c}
~~; \\
\label{eq:aMchi}
&& \sA_{\rm G}=\frac{a_{\rm G}}{(a_{\rm G})_{\rm c}}~~;\qquad
\sM_{\rm G}=\frac{M_{\rm G}}{(M_{\rm G})_{\rm c}}~~;\qquad
\chi_{\rm c}=\frac{1+(\nu_{\rm GP})_{\rm tid}/(\nu_{\rm G})_{\rm sel}}
{1+[(\nu_{\rm GP})_{\rm tid}]_{\rm c}/(\nu_{\rm G})_{\rm sel}}~~;\qquad
\end{lefteqnarray}
where $\sA_{\rm G}$ and $\sM_{\rm G}$ may be conceived as reduced radius and
reduced mass, respectively.

\begin{figure*}[t]
\begin{center}
\includegraphics[scale=0.8]{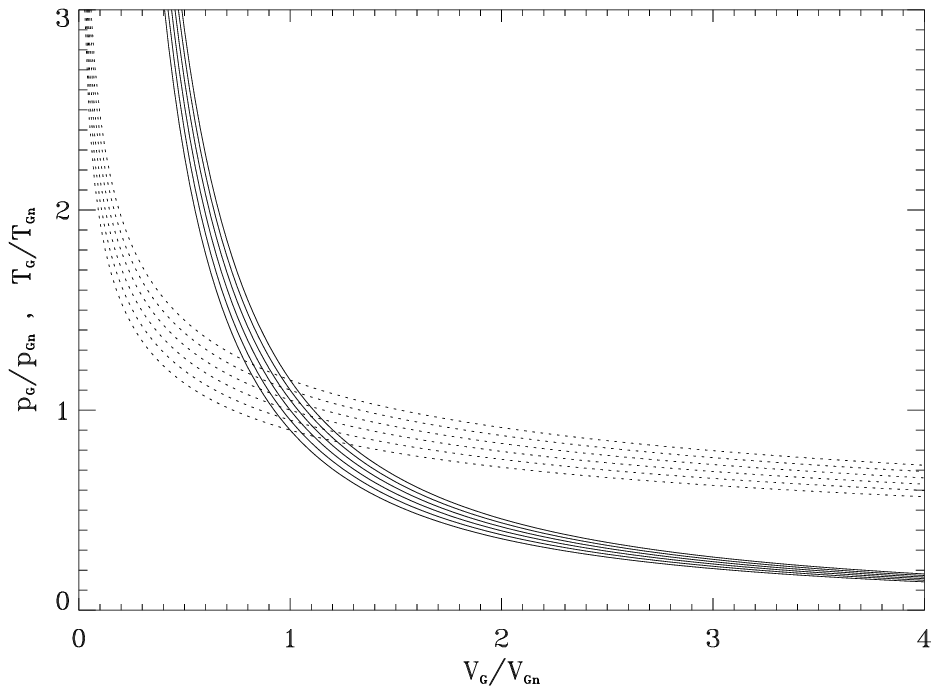}
\caption{Reduced macropressure, $\sP_{\rm G}$, and reduced macrotemperature,
$\sT_{\rm G}$, vs reduced volume, $\sV_{\rm G}$, for
$\kappa=0.90, 0.95, 1.00, 1.05, 1.10, 1.15$, from
bottom to top, where $\kappa=M_{\rm G}^2$ for reduced macropressure (full
curves) and $\kappa=M_{\rm G}\chi_{\rm c}$ for reduced macrotemperature
(dotted curves).}
\label{f:pVT101}
\end{center}
\end{figure*}

Reduced macropressure and reduced macrotemperature, as functions of reduced
macrovolume, read:
\begin{lefteqnarray}
\label{eq:pTVc}  
&& \sP_{\rm G}=\sM_{\rm G}^2\sV_{\rm G}^{-4/3}~~;\qquad\sT_{\rm G}=
\sM_{\rm G}\sV_{\rm G}^{-1/3}\chi_{\rm c}~~;
\end{lefteqnarray}
where the dependence on the path along a fractional isoenergetic,
$\phi={\rm const}$, occurs via $\chi_{\rm c}$.   The dependence of reduced
macropressure, $\sP_{\rm G}$, and reduced macrotemperature, $\sT_{\rm G}$, on
reduced macrovolume, $\sV_{\rm G}$, for assigned reduced mass, $\sM_{\rm G}$,
and parameter, $\chi_{\rm c}$, is shown in Fig.\,\ref{f:pVT101} where macrogas
state is defined by the intersection of two selected curves together with
related point on fractional isoenergetic, $(m,y,\phi)$.

The dependence of reduced macrotemperature i.e. $\chi_{\rm c}$, on reduced
coordinate, $\sX_{\rm V}$, for fixed reduced volume and reduced mass, or
reduced coordinate, $\sX_{\rm T}$, is owing to density profiles and shall be
determined for UU, HH, HN/NH macrogases in the following sections.  

\subsection{UU macrogases}\label{UU}

UU macrogases exhibit flat density profiles, which is equivalent to polytropes
of index, $n=0$ (e.g., Chandrasekhar 1939, Chap.\,IV, \S4; Caimmi 1986; Caimmi
2016), but implies negative distribution functions for stellar fluids
(Vandervoort 1980).   For this reason, UU macrogases are of little
astrophysical interest and can be considered as a simple guidance case.

Fractional isoenergetics $(X_{\rm T}={\rm const})$ may be explicitly
expressed as (C10):
\begin{leftsubeqnarray}
\slabel{eq:UUXpa}
&& X_{\rm p}=\frac{X_{\rm T}}{X_{\rm V}}[1+\Psi(X_{\rm V},X_{\rm T})]~~; \\
\slabel{eq:UUXpb}
&& \Psi(X_{\rm V},X_{\rm T})=
\cases{
\frac{X_{\rm V}^5\Phi^2(X_{\rm V},X_{\rm T})}{2X_{\rm T}}-  
\frac{X_{\rm V}^3\Phi(X_{\rm V},X_{\rm T})}{X_{\rm T}}        
\sqrt{\frac{X_{\rm V}^4\Phi^2(X_{\rm V},X_{\rm T})}4+\frac{X_{\rm T}}
{X_{\rm V}}}~~;                                                          & \cr
0<X_{\rm V}\le1~~;                                                       & \cr
\frac{\Phi^2(X_{\rm V},X_{\rm T})}{2X_{\rm V}^5X_{\rm T}}-\frac
{\Phi(X_{\rm V},X_{\rm T})}{X_{\rm V}^2X_{\rm T}}
\sqrt{\frac{\Phi^2(X_{\rm V},X_{\rm T})}{4X_{\rm V}^6}+\frac
{X_{\rm T}}{X_{\rm V}}}~~;                                               & \cr
1\le X_{\rm V}<+\infty~~;                                                & \cr
} \\
\slabel{eq:UUXpc}
&& \Phi(X_{\rm V},X_{\rm T})=
\cases{
\frac52\frac1{X_{\rm V}^2}-\frac32-X_{\rm T}~~; &
$0<X_{\rm V}\le1~~;$                                                       \cr
1-\left(\frac52X_{\rm V}^2-\frac32\right)X_{\rm T}~~; &                  
$1\le X_{\rm V}<+\infty~~;$                                                \cr
} \\
\slabel{eq:UUXpd}
&& \lim_{X_{\rm V}\to0^+}X_{\rm p}=\lim_{X_{\rm V}\to0^+}\frac{X_{\rm T}}
{X_{\rm V}}=+\infty~~;\quad
   \lim_{X_{\rm V}\to+\infty}X_{\rm p}=\lim_{X_{\rm V}\to+\infty}\frac
{X_{\rm T}}{X_{\rm V}}=0~~;\qquad \\
\slabel{eq:UUXpe}
&& \lim_{X_{\rm T}\to0^+}X_{\rm p}=0~~;\qquad
   \lim_{X_{\rm T}\to+\infty}X_{\rm p}=+\infty~~;\qquad
\label{seq:UUXp}
\end{leftsubeqnarray}
where $X_{\rm p}$, $X_{\rm V}$, $X_{\rm T}$, are dimensionless variables
defined by Eq.\,(\ref{eq:XpVT}).

Plotting fractional isoenergetics, $X_{\rm p}(X_{\rm V})$, for
assigned $X_{\rm T}$, on $({\sf O}X_{\rm V}X_{\rm p})$ plane, shows two
extremum points (minimum and maximum, respectively) but no critical point
within the domain, $0<X_{\rm T}<+\infty$ (C10).
On the other hand, $\phi_{\rm crit}\approx10$, $m_{\rm crit}\approx10$, for
HH and HN/NH macrogases (C10).   Then $\phi_{\rm norm}=10$ shall be
arbitrarily assumed as normalization value, together with
$y_{\rm norm}=1.186944$ implying $m_{\rm norm}=7.290778$ via
Eqs.\,(\ref{eq:UUXpa})-(\ref{eq:UUXpc}).   The result is:
\begin{lefteqnarray}
\label{eq:UUnr}
&& X_{\rm V_n}=\frac1{y_{\rm norm}}=0.8425~~;\quad X_{\rm p_n}=
(m_{\rm norm})^2=53.15545~~;\nonumber \\
&&  X_{\rm T_n}=\phi_{\rm norm}=10~~;\qquad
\end{lefteqnarray}
and reduced variables:
\begin{lefteqnarray}
\label{eq:UUrd}
&& \sX_{\rm V}=\frac{X_{\rm V}}{X_{\rm V_n}}~~;\qquad
   \sX_{\rm p}=\frac{X_{\rm p}}{X_{\rm p_n}}~~;\qquad
   \sX_{\rm T}=\frac{X_{\rm T}}{X_{\rm T_n}}~~;\qquad
\end{lefteqnarray}
shall be used in place of their counterparts related to critical values.

Fractional isoenergetics, $\sX_{\rm p}^{-1}(\sX_{\rm V}^{-1})$, are plotted in
Fig.\,\ref{f:uusp} for
$\sX_{\rm T}^{-1}=20/23$, $20/22, 20/21, 20/20, 20/19, 20/18$, from bottom to
top. The contribution from the term, $\sX_{\rm T}^{-1}/\sX_{\rm V}^{-1}$,
is shown by dotted curves.   More specifically, dividing both sides of
Eq.\,(\ref{eq:UUXpa}) by their counterparts related to
$(X_{\rm Vn}, X_{\rm pn},X_{\rm Tn})$ yields:
\begin{figure*}[t]
\begin{center}
\includegraphics[scale=0.8]{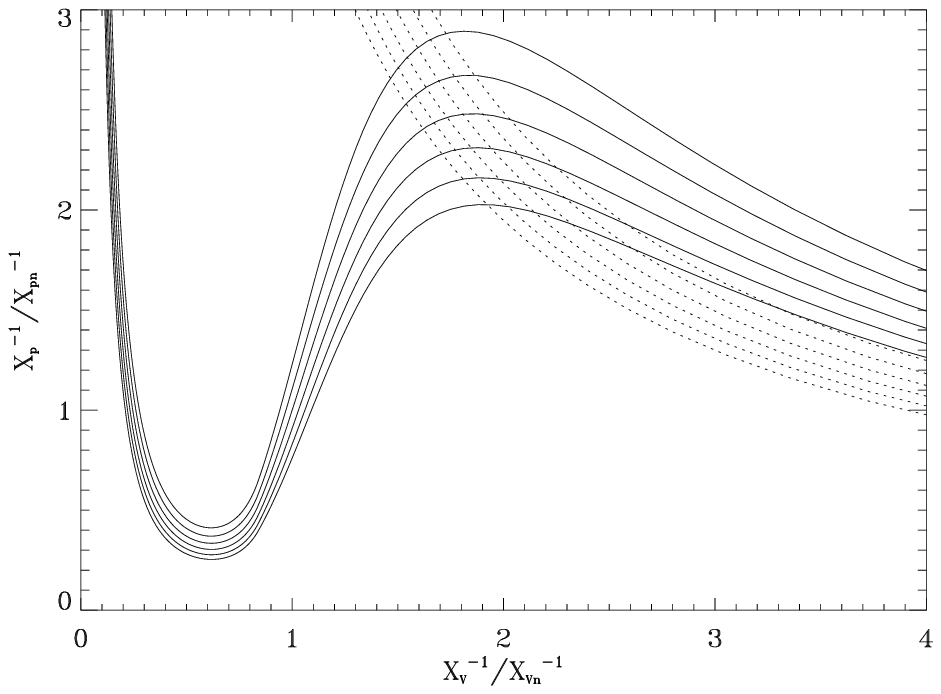}
\caption{Fractional isoenergetics related to UU macrogases for 
$\sX_{\rm T}^{-1}=20/23, 20/22, 20/21, 20/20, 20/19, 20/18$, from bottom to
top.   The contribution from the term, $\sX_{\rm T}^{-1}/\sX_{\rm V}^{-1}$,
is shown by dotted curves.   Normalization values are
$(X_{\rm Vn}, X_{\rm pn},X_{\rm Tn})=(0.8425,53.15545,10)$.}
\label{f:uusp}
\end{center}
\end{figure*}
\begin{lefteqnarray}
\label{eq:fuui}  
&& \sX_{\rm p}=\frac{\sX_{\rm T}}{\sX_{\rm V}}
\frac{1+\Psi(X_{\rm V},X_{\rm T})}{1+\Psi(X_{\rm Vn},X_{\rm Tn})}~~;
\end{lefteqnarray}
where the inverse of the whole product and the first factor on the right-hand
side are
represented in Fig.\,\ref{f:uusp} as full and dotted curves, respectively.
The reason for plotting $\sX_{\rm p}^{-1}$ vs $\sX_{\rm V}^{-1}$ instead of
$\sX_{\rm p}$ vs $\sX_{\rm V}$ shall be clarified in dealing with HH and HN/NH
macrogases.

In the case under consideration of flat density profiles, without loss of
generality, scaled truncation radii can be assumed as
$\Xi_{\rm U}=R_{\rm U}/r_{\rm U}^\dagger=1$, U = G,P.   Accordingly, profile
factors reduce to:
\begin{lefteqnarray}
\label{eq:Unum}  
&& (\nu_{\rm U})_{\rm mas}=1~~;\qquad(\nu_{\rm U})_{\rm sel}=\frac3{10}~~;
\qquad{\rm U=G,P}~~; \\
&& (\nu_{\rm GP})_{\rm tid}=
\cases{
\frac3{10}\frac m{y^3}~~;                            & $1\le y<+\infty~~;$ \cr
\frac3{10}m\left(\frac52-\frac32y^2\right)~~;        & $0<y\le1~~;$        \cr
} \\
&& (\nu_{\rm PG})_{\rm tid}=
\cases{
\frac3{10}\frac1m\left(\frac52-\frac32\frac1{y^2}\right)~~;
                                                  & $1\le y<+\infty~~;~~;$ \cr
\frac3{10}\frac{y^3} m~~;                         & $0<y\le1~~;$           \cr
}
\end{lefteqnarray}
for further details, an interested reader is addressed to the parent paper
(C10).   The special case of coinciding volumes, $y=1$, implies
$(\nu_{\rm GP})_{\rm tid}=(3/10)m$; $(\nu_{\rm PG})_{\rm tid}=(3/10)(1/m)$;
as expected.

In addition, Eqs.\,(\ref{eq:virFb}) and (\ref{eq:aMchi}) reduce to:
\begin{lefteqnarray}
\label{eq:UFGP}  
&& F_{\rm GP}=
\cases{
\frac3{10}\left(1+\frac m{y^3}\right)\Sigma~~;       & $1\le y<+\infty~~;$ \cr
\frac3{10}\left[1+m\left(\frac52-\frac32y^2\right)\right]\Sigma~~; 
                                                     & $0<y\le1~~;$        \cr
} \\
\label{eq:UFPG}  
&& F_{\rm PG}=
\cases{
\frac3{10}\left[1+\frac1m\left(\frac52-\frac32\frac1{y^2}\right)\right]\Sigma
~~;                                                  & $1\le y<+\infty~~;$ \cr
\frac3{10}\left(1+\frac{y^3}m\right)\Sigma~~;        & $0<y\le1~~;$        \cr
} \\
\label{eq:Uchi}  
&& \chi_{\rm c}=
\cases{                                                           
\frac{1+m/y^3}{1+m_{\rm norm}/y_{\rm norm}^3}~~; & $1\le y<+\infty~~;$     \cr
\frac{1+m[5/2-(3/2)y^2]}{1+m_{\rm norm}/y_{\rm norm}^3}~~;
                                                 & $0<y\le1~~;$            \cr
}
\end{lefteqnarray}
where $y=a_{\rm P}/a_{\rm G}$.

\begin{figure*}[t]
\begin{center}
\includegraphics[scale=0.8]{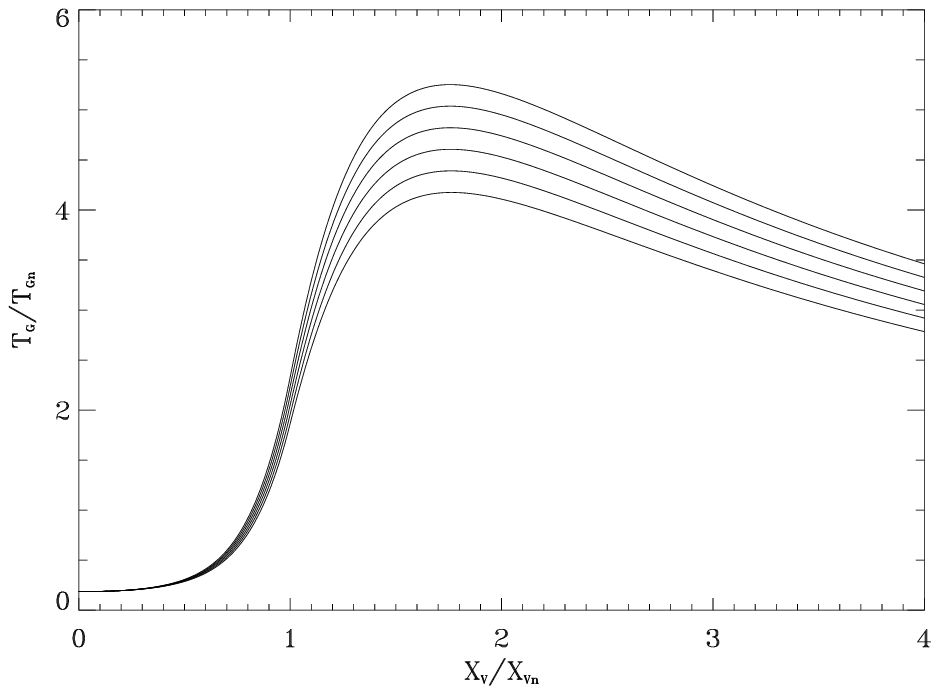}
\caption{Reduced macrotemperature, $\sT_{\rm G}$, vs reduced variable,
$\sX_{\rm V}$, related to UU macrogases for
$\sM_{\rm G}^{-1}\sV_{\rm G}^{1/3}=1$ and
$\sX_{\rm T}^{-1}=20/23,20/22,20/21$, $20/20,20/19,20/18$,
from bottom to top.
A similar trend is exhibited for critical fractional isoenergetic,
$\sX_{\rm T}^{-1}=1$, and
$\sM_{\rm G}^{-1}\sV_{\rm G}^{1/3}=20/23,20/22,20/21$, $20/20,20/19,20/18$,
from bottom to top.}
\label{f:uuTchi}
\end{center}
\end{figure*}

The dependence of reduced macrotemperature, $\sT_{\rm G}$, on reduced
variable, $\sX_{\rm V}$, is shown in Fig.\,\ref{f:uuTchi} for
$\sM_{\rm G}^{-1}\sV_{\rm G}^{1/3}=1$ and
$\sX_{\rm T}^{-1}=20/23,20/22,20/21$, 20/20, 20/19, 20/18, from bottom to top.
A similar trend is exhibited for critical fractional isoenergetic,
$\sX_{\rm T}^{-1}=1$, and
$\sM_{\rm G}^{-1}\sV_{\rm G}^{1/3}=20/23,20/22,20/21$, $20/20, 20/19, 20/18$,
from bottom to top.

\subsection{HH macrogases}\label{HH}

HH macrogases exhibit central cusp and vanishing density at infinite distance
(Hernquist 1990), and have been proved to be consistent with nonnegative
distribution functions in an acceptable parameter range (Ciotti 1996).

Fractional isoenergetics on $({\sf O}\sX_{\rm V}\sX_{\rm p})$ plane show weak
dependence on scaled truncation radii, $\Xi_{\rm G}$, $\Xi_{\rm P}$, including
the limit, $\Xi_{\rm G}\to+\infty$, $\Xi_{\rm P}\to+\infty$, where formulation
is far simpler.   For this reason, attention shall be restricted to infinite
scaled truncation radii, which imply infinitesimal scaling radii,
$r_{\rm U}^\dagger$, and/or infinite truncation radii, $R_{\rm U}$, U = G,P,
along an arbitrary direction.   Accordingly, profile factors reduce to (C10):
\begin{lefteqnarray}
\label{eq:nusH}
&& (\nu_{\rm U})_{\rm mas}=12~~;\quad(\nu_{\rm U})_{\rm sel}=12~~;\quad
\Xi_{\rm U}\to+\infty~~;\quad{\rm U=G,P}~~;\quad \\
\label{eq:nuHGP}
&& (\nu_{\rm GP})_{\rm tid}=
\cases{
-\frac98mw^{(\rm ext)}(z)~~;        &             $1\le y<+\infty~~;$      \cr
-\frac98\frac myw^{(\rm int)}(z)~~; &             $0<y\le1~~;$             \cr
} \\
\label{eq:nuHPG}
&& (\nu_{\rm PG})_{\rm tid}=                      
\cases{                                          
-\frac98\frac ymw^{(\rm int)}(z)~~; &             $1\le y<+\infty~~;$      \cr
-\frac98\frac1mw^{(\rm ext)}(z)~~;  &             $0<y\le1~~;$             \cr
} \\
\label{eq:wiH}
&& w^{({\rm int})}(z)=
\cases{
-\frac{64z}{(z-1)^4}\left[-2(2z+1)\ln z+(z-1)(z+5)\right]~~;  & $z\ne1~~;$ \cr
-\frac{32}3~~;\qquad                                          & $z=1~~;$   \cr
 } \\
\label{eq:weH}
&& w^{({\rm ext})}(z)=
\cases{
-\frac{64}{(z-1)^4}\left[2z(z+2)\ln z-(z-1)(5z+1)\right]~~;   & $z\ne1~~;$ \cr
-\frac{32}3~~;                                                & $z=1~~;$   \cr
} \\
\label{eq:zH}
&& z=y^\dagger=\frac{\Xi_{\rm G}}{\Xi_{\rm P}}y~~;
\end{lefteqnarray}
where $y=a_{\rm P}/a_{\rm G}$.   For further details, an interested reader is
addressed to the parent paper (C10) and an earlier investigation (Caimmi and
Valentinuzzi 2008).

\begin{figure*}[t]
\begin{center}
\includegraphics[scale=0.8]{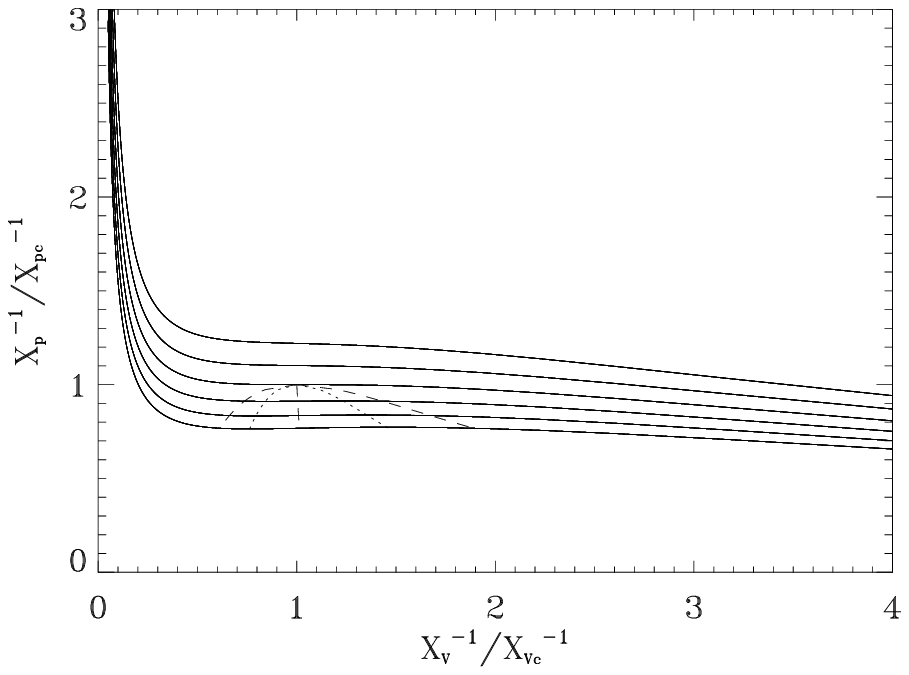}
\caption{Fractional isoenergetics related to HH macrogases for
$\sX_{\rm T}^{-1}=20/23,20/22,20/21,20/20,20/19,20/18$, from bottom to top,
where cases, $\Xi_{\rm G}/\Xi_{\rm P}=0.25$, 0.50, 1.00, 2.00, 4.00, are
superimposed.   The dashed curve (including central branch) is the locus of
intersections between HH fractional isoenergetics and horizontal lines
yielding
regions of equal area.   The dotted curve is the locus of HH fractional
isoenergetic extremum points.   Plotting $\sX_{\rm p}^{-1}$ vs
$\sX_{\rm V}^{-1}$ yields fractional isoenergetics similar to VDW 
isothermals shown in Fig.\,\ref{f:vris100}, where extremum points are lying
below the critical point.}
\label{f:hhsp}
\end{center}
\end{figure*}

Fractional isoenergetics $(\sX_{\rm T}^{-1}={\rm const})$, explicitly
expressed
substituting Eqs.\,(\ref{eq:virFb}) and (\ref{eq:nusH})-(\ref{eq:zH}) into
(\ref{eq:eqse}), are plotted in Fig.\,\ref{f:hhsp} for
$\sX_{\rm T}^{-1}=20/23,20/22$, $20/21,20/20,20/19,20/18$, from bottom to top,
where cases, $\Xi_{\rm G}/\Xi_{\rm P}=0.25$, 0.50, 1.00, 2.00, 4.00, are
superimposed.
The dashed curve (including central branch) is the locus of intersections
between HH fractional isoenergetics and horizontal lines yielding regions of
equal area.   The dotted curve is the locus of HH fractional isoenergetic
extremum points.   Plotting $\sX_{\rm p}^{-1}$ vs
$\sX_{\rm V}^{-1}$ yields fractional isoenergetics similar to VDW 
isothermals shown in Fig.\,\ref{f:vris100}, where extremum points are lying
below the horizontal inflexion point.

Critical points, $(X_{\rm Vc}, X_{\rm pc},$ $X_{\rm Tc})$, can be inferred
from Table\,\ref{t:hhcri} via Eq.\,(\ref{eq:XpVT}).
\begin{table}
\caption{Values of scaling fractional
mass, $m^\dagger=M(r_{\rm P}^\dagger)/M(r_{\rm G}^\dagger)$, fractional mass,
$m$, scaling fractional radius, $y^\dagger$, 
fractional truncation radius, $y$, and fractional
energy, $\phi$, related to
critical point i.e. the horizontal inflexion
point on the critical fractional isoenergetic,
for selected fractional scaled truncation radii,
$\Xi_{\rm G}/\Xi_{\rm P}$, with regard to HH density profiles.}
\label{t:hhcri}
\begin{center}
\begin{tabular}{|c|c|c|c|c|c|} \hline
$\Xi_{\rm G}/\Xi_{\rm P}$ & $m^\dagger$ & $m$ & $y^\dagger$ & $y$ & $\varphi$
\\
\hline
0.25 & 20.2157 & 20.2157 & 4.2153 & 16.8612 & 18.1509 \\
0.50 & 20.2148 & 20.2148 & 4.2594 & 08.5188 & 18.1500 \\
1.00 & 20.2148 & 20.2148 & 4.2594 & 04.2594 & 18.1500 \\
2.00 & 20.2148 & 20.2148 & 4.2641 & 02.1320 & 18.1500 \\
4.00 & 20.2148 & 20.2148 & 4.2656 & 01.0664 & 18.1500 \\
\hline
\end{tabular}                                                                                       
\end{center}                                                                                        
\end{table}                                                                                         
More specifically, values of scaling fractional
mass, $m^\dagger=M(r_{\rm P}^\dagger)/M(r_{\rm G}^\dagger)$, fractional mass,
$m$, fractional scaling radius, $y^\dagger$, 
fractional truncation radius, $y$, and fractional
energy, $\phi$, related to
critical point i.e. the horizontal inflexion
point on the critical fractional isoenergetic,
are listed in Table\,\ref{t:hhcri} for selected fractional scaled
truncation radii,
$\Xi_{\rm G}/\Xi_{\rm P}$, with regard to HH density profiles.

An inspection of Table\,\ref{t:hhcri} discloses weak dependence of critical
parameters on the ratio, $\Xi_{\rm G}/\Xi_{\rm P}$, within the range
considered, leaving aside $y$ via Eq.\,(\ref{eq:zH}).   Accordingly, critical
parameters might be conceived as independent of (infinite) scaled truncation
radii to a first extent.

An inspection of Fig.\,\ref{f:hhsp} discloses weak dependence of fractional
isoenergetics on  the ratio, $\Xi_{\rm G}/\Xi_{\rm P}$, within the range
considered.   Accordingly, fractional isoenergetics might be conceived as
independent of (infinite) scaled truncation radii to a first extent.

\begin{figure*}[t]
\begin{center}
\includegraphics[scale=0.8]{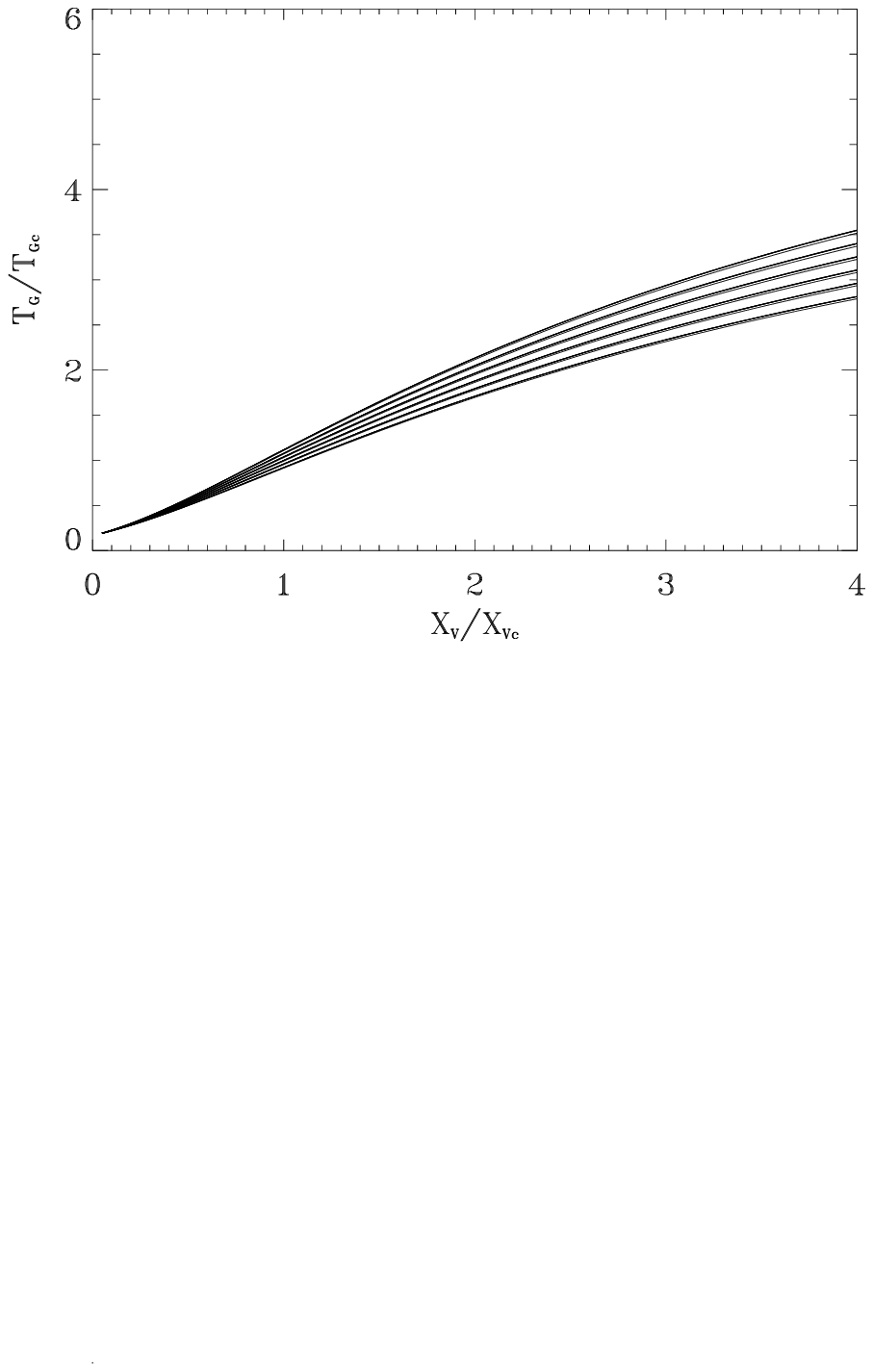}
\caption{Fractional macrotemperature, $\sT_{\rm G}$, vs reduced variable,
$\sX_{\rm V}$, related to HH macrogases for
$\sM_{\rm G}^{-1}\sV_{\rm G}^{1/3}=1$ and
$\sX_{\rm T}^{-1}=20/23,20/22,20/21$, $20/20,20/19,20/18$, from bottom to top,
where cases, $\Xi_{\rm G}/\Xi_{\rm P}=0.25, 0.50$, $1.00, 2.00, 4.00$, are
superimposed.  A similar trend is exhibited for critical fractional
isoenergetic, $\sX_{\rm T}^{-1}=1$, and
$\sM_{\rm G}^{-1}\sV_{\rm G}^{1/3}=20/23,20/22,20/21$, $20/20,20/19,20/18$,
from bottom to top.}
\label{f:hhTchi}
\end{center}
\end{figure*}

The dependence of reduced macrotemperature, $\sT_{\rm G}$, on reduced
variable, $\sX_{\rm V}$, is shown in Fig.\,\ref{f:hhTchi} for
$\sM_{\rm G}^{-1}\sV_{\rm G}^{1/3}=1$ and
$\sX_{\rm T}^{-1}=20/23,20/22,20/21$, $20/20,20/19,20/18$, from bottom to top.
A similar trend is exhibited for critical fractional isoenergetic,
$\sX_{\rm T}^{-1}=1$, and
$\sM_{\rm G}^{-1}\sV_{\rm G}^{1/3}=20/23,20/22,20/21$, $20/20,20/19,20/18$,
from bottom to top.

An inspection of Fig.\,\ref{f:hhTchi} discloses weak dependence of reduced
macrotemperature on  the ratio, $\Xi_{\rm G}/\Xi_{\rm P}$, within the range
considered.   Accordingly, reduced macrotemperature might be conceived as
independent of (infinite) scaled truncation radii to a first extent.

\subsection{HN/NH macrogases}\label{HN}

HN/NH macrogases exhibit central cusp and vanishing density at infinite
distance, where HN means inner H density profile, related to G macrogas, and
outer N density profile, related to P macrogas, $y=a_{\rm P}/a_{\rm G}\ge1$,
and NH means inner N and outer H, $y=a_{\rm P}/a_{\rm G}\le1$.   N density
profiles decline more slowly with respect to above considered H, yielding
infinite mass within infinite radius (Navarro et al. 1995, 1996, 1997).   HN
macrogases have been proved to be consistent with nonnegative distribution
functions in an acceptable parameter range (Lowenstein and White 1999) by use
of a previously stated theorem (Ciotti and Pellegrini 1992).

Fractional isoenergetics on $({\sf O}\sX_{\rm V}\sX_{\rm p})$ plane show weak
dependence on scaled truncation radii, $\Xi_{\rm G}$, $\Xi_{\rm P}$, including
the limit, $\Xi_{\rm G}\to+\infty$, $\Xi_{\rm P}\to+\infty$, where formulation
is far simpler.   For this reason, attention shall be restricted to infinite
scaled truncation radii, which imply infinitesimal scaling radii,
$r_{\rm U}^\dagger$, and/or infinite truncation radii, $R_{\rm U}$, U = G,P,
along an arbitrary direction.   Accordingly, profile factors reduce to (C10):
\begin{lefteqnarray}
\label{eq:nusN}
&& (\nu_{\rm U})_{\rm mas}\to+\infty~~;\quad(\nu_{\rm U})_{\rm sel}=36~~;\quad
\Xi_{\rm U}\to+\infty~~; \\
&& {\rm U}=\cases{
{\rm G}~~; & NH macrogases \cr
{\rm P}~~; & HN macrogases \cr
} \nonumber \\
\label{eq:nuNGP}
&& (\nu_{\rm GP})_{\rm tid}=
\cases{
-\frac98m^\dagger w_{\rm HN}^{(\rm ext)}(z)~~;           & $1\le y<+\infty~~;$
\cr
-\frac98\frac{m^\dagger}{y^\dagger}w_{\rm NH}^{(\rm int)}(z)~~; & $0<y\le1~~;$
\cr
} \\
\label{eq:nuNPG}
&& (\nu_{\rm PG})_{\rm tid}=                      
\cases{                                          
-\frac98\frac{y^\dagger}{m^\dagger}w_{\rm HN}^{(\rm int)}(z)~~; &
$1\le y<+\infty~~;$ \cr
-\frac98\frac1{m^\dagger}w_{\rm NH}^{(\rm ext)}(z)~~;           & $0<y\le1~~;$
\cr
} \\
\label{eq:wiHN}
&& w_{\rm HN}^{({\rm int})}(z)=
\cases{
-\frac{64}{(z-1)^3}(-2z\ln z+z^2-1)~~; & $z\ne1~~;$ \cr
-\frac{64}3~~;                         & $z=1~~;$   \cr
 } \\
\label{eq:weHN}
&& w_{\rm HN}^{({\rm ext})}(z)=
\cases{
-\frac{64}{(z-1)^2}\left(\frac{z+1}{z-1}\ln z-2\right)~~; & $z\ne1~~;$ \cr
-\frac{32}3~~;                                            & $z=1~~;$   \cr
} \\
\label{eq:wiNH}
&& w_{\rm NH}^{({\rm int})}(z)=
\cases{
-\frac{64z}{(z-1)^2}\left(\frac{z+1}{z-1}\ln z-2\right)~~;  & $z\ne1~~;$ \cr
-\frac{32}3~~;\qquad                                        & $z=1~~;$   \cr
 } \\
\label{eq:weNH}
&& w_{\rm NH}^{({\rm ext})}(z)=
\cases{
-\frac{64}{(z-1)^3}(z^2-1-2z\ln z)~~; & $z\ne1~~;$ \cr
-\frac{64}3~~;                        & $z=1~~;$   \cr
} \\
\label{eq:zN}
&& z=y^\dagger=\frac{\Xi_{\rm G}}{\Xi_{\rm P}}y~~;
\end{lefteqnarray}
where $y=a_{\rm P}/a_{\rm G}$.   For further details, an interested reader is
addressed to the parent paper (C10) and an earlier investigation (Caimmi and
Valentinuzzi 2008).

\begin{figure*}[t]
\begin{center}
\includegraphics[scale=0.8]{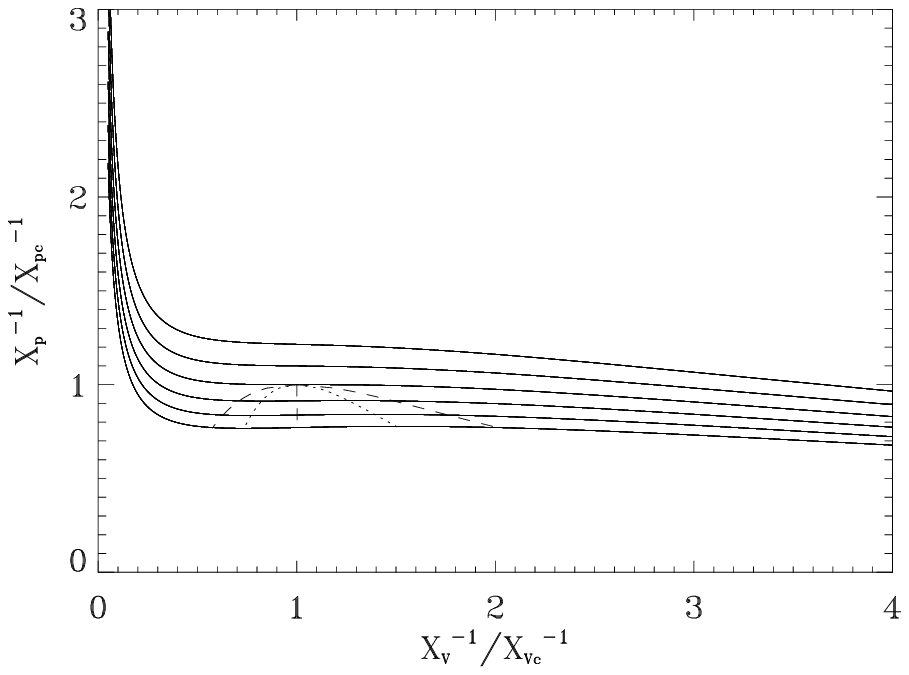}
\caption{Fractional isoenergetics related to HN/NH macrogases for
$\sX_{\rm T}^{-1}=20/23, 20/22, 20/21, 20/20, 20/19, 20/18$, from bottom to
top, where cases, $\Xi_{\rm G}/\Xi_{\rm P}=0.25$, 0.50, 1.00, 2.00, 4.00, are
superimposed.   The dashed curve (including central branch) is the locus of
intersections between HN/NH fractional isoenergetics and horizontal lines
yielding regions of equal area.   The dotted curve is the locus of HN/NH
fractional isoenergetic extremum points.   Plotting $\sX_{\rm p}^{-1}$ vs
$\sX_{\rm V}^{-1}$ yields fractional isoenergetics similar to VDW 
isothermals shown in Fig.\,\ref{f:vris100}, where extremum points are lying
below the critical point.}
\label{f:hnsp}
\end{center}
\end{figure*}

Fractional isoenergetics $(\sX_{\rm T}^{-1}={\rm const})$, explicitly
expressed
substituting Eqs.\,(\ref{eq:virFb}) and (\ref{eq:nusN})-(\ref{eq:zN}) into
(\ref{eq:eqse}), are plotted in Fig.\,\ref{f:hnsp} for
$\sX_{\rm T}^{-1}=20/23$, $20/22, 20/21, 20/20, 20/19,20/18$, from bottom to
top, where cases, $\Xi_{\rm G}/\Xi_{\rm P}=0.25$, 0.50, 1.00, 2.00, 4.00, are
superimposed.
The dashed curve (including central branch) is the locus of intersections
between HN/NH fractional isoenergetics and horizontal lines yielding regions of
equal area.   The dotted curve is the locus of HN/NH fractional isoenergetic
extremum points.   Plotting $\sX_{\rm p}^{-1}$ vs
$\sX_{\rm V}^{-1}$ yields fractional isoenergetics similar to VDW
isothermals shown in Fig.\,\ref{f:vris100}, where extremum points are lying
below the horizontal inflexion point.

Critical points, $(X_{\rm Vc}, X_{\rm pc},$ $X_{\rm Tc})$, can be inferred
from Table\,\ref{t:hncri} via Eq.\,(\ref{eq:XpVT}).
\begin{table}
\caption{Values of scaling fractional
mass, $m^\dagger=M(r_{\rm P}^\dagger)/M(r_{\rm G}^\dagger)$, fractional mass,
$m$, scaling fractional radius, $y^\dagger$, 
fractional truncation radius, $y$, and fractional
energy, $\phi$, related to
critical point i.e. the horizontal inflexion
point on the critical fractional isoenergetic,
for selected fractional scaled truncation radii,
$\Xi_{\rm G}/\Xi_{\rm P}$, with regard to HN/NH density profiles.}
\label{t:hncri}
\begin{center}
\begin{tabular}{|c|c|c|c|c|c|} \hline
$\Xi_{\rm G}/\Xi_{\rm P}$ & $m^\dagger$ & $m$ & $y^\dagger$ & $y$ & $\varphi$
\\
\hline
0.25 & 12.4148 & $\infty$ & 1.9785 & 7.9142 & 35.8702 \\
0.50 & 12.4006 & $\infty$ & 2.0149 & 4.0298 & 35.8259 \\
1.00 & 12.3984 & $\infty$ & 2.0292 & 2.0292 & 35.8190 \\
2.00 & 12.3984 & $\infty$ & 2.0303 & 1.0151 & 35.8190 \\
4.00 & 12.3984 & $\infty$ & 2.0303 & 0.5076 & 35.8190 \\
\hline
\end{tabular}                                                                                       
\end{center}                                                                                        
\end{table}                                                                                         
More specifically, values of scaling fractional
mass, $m^\dagger=M(r_{\rm P}^\dagger)/M(r_{\rm G}^\dagger)$, fractional mass,
$m$, fractional scaling radius, $y^\dagger$, 
fractional truncation radius, $y$, and fractional
energy, $\phi$, related to
critical point i.e. the horizontal inflexion
point on the critical fractional isoenergetic,
are listed in Table\,\ref{t:hncri} for selected fractional scaled
truncation radii,
$\Xi_{\rm G}/\Xi_{\rm P}$, with regard to HN/NH density profiles.

An inspection of Table\,\ref{t:hncri} discloses weak dependence of critical
parameters on the ratio, $\Xi_{\rm G}/\Xi_{\rm P}$, within the range
considered, leaving aside $y$ via Eq.\,(\ref{eq:zN}).   Accordingly, critical
parameters might be conceived as independent of (infinite) scaled truncation
radii to a first extent.

An inspection of Fig.\,\ref{f:hnsp} discloses weak dependence of fractional
isoenergetics on  the ratio, $\Xi_{\rm G}/\Xi_{\rm P}$, within the range
considered.   Accordingly, fractional isoenergetics might be conceived as
independent of (infinite) scaled truncation radii to a first extent.

\begin{figure*}[t]
\begin{center}
\includegraphics[scale=0.8]{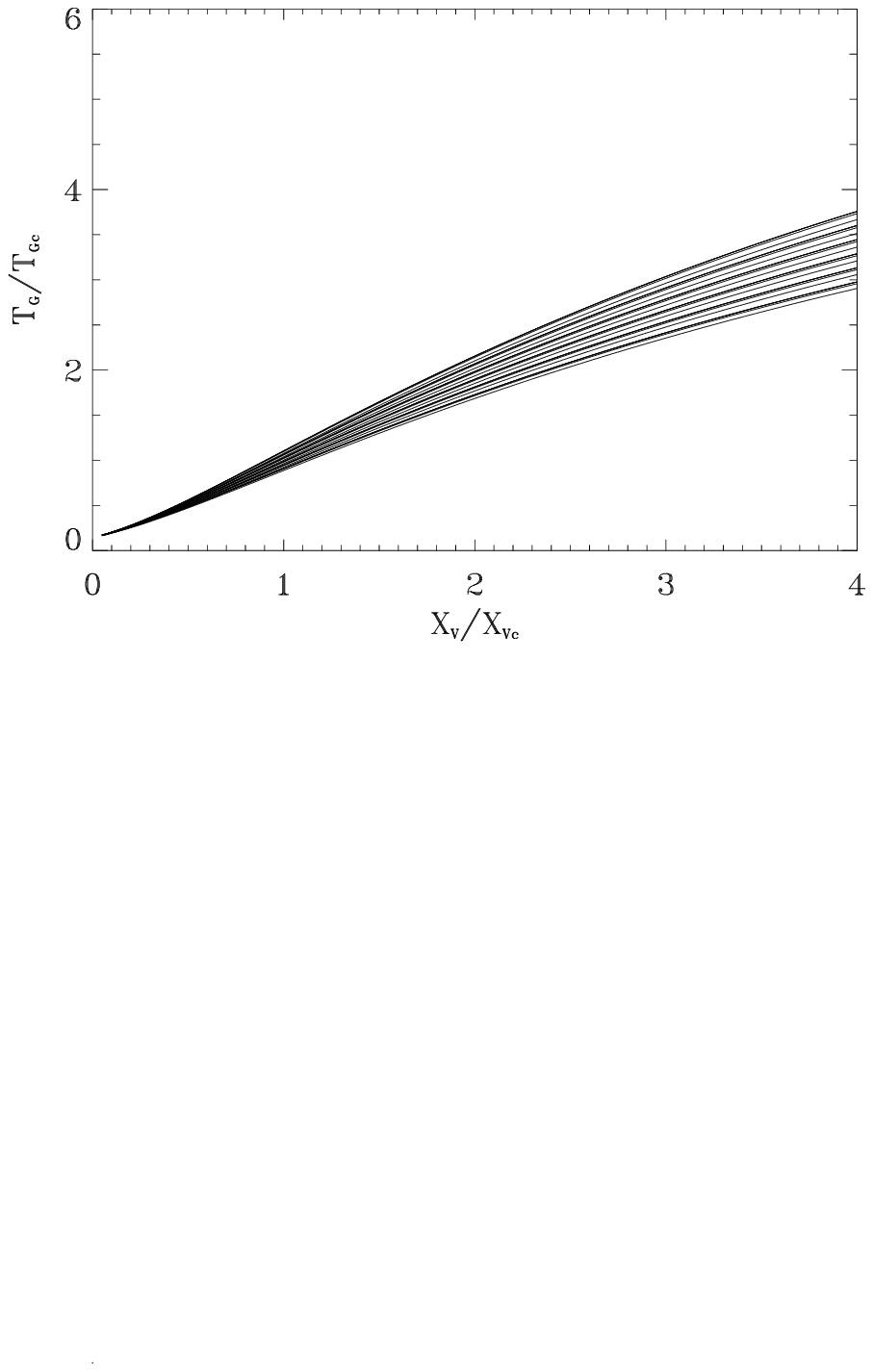}
\caption{Reduced macrotemperature, $\sT_{\rm G}$, vs reduced variable,
$\sX_{\rm V}$, related to HN/NH macrogases for
$\sM_{\rm G}^{-1}\sV_{\rm G}^{1/3}=1$
and $\sX_{\rm T}^{-1}=20/23, 20/22$, $20/21, 20/20, 20/19, 20/18$, from bottom
to top,
where cases, $\Xi_{\rm G}/\Xi_{\rm P}=0.25$, $0.50, 1.00, 2.00, 4.00$, are
superimposed.  A similar trend is exhibited for critical fractional
isoenergetic,
$\sX_{\rm T}=1$, and $\sM_{\rm G}^{-1}\sV_{\rm G}^{1/3}=20/23, 20/22, 20/21$,
$20/20, 20/19, 20/18$, from bottom to top.}
\label{f:hnTchi}
\end{center}
\end{figure*}

The dependence of reduced macrotemperature, $\sT_{\rm G}$, on reduced
variable, $\sX_{\rm V}$, is shown in Fig.\,\ref{f:hnTchi} for
$\sM_{\rm G}^{-1}\sV_{\rm G}^{1/3}=1$ and
$\sX_{\rm T}^{-1}=20/23, 20/22, 20/21$, $20/20, 20/19, 20/18$, from bottom to
top.
A similar trend is exhibited for critical fractional isoenergetic,
$\sX_{\rm T}^{-1}=1$, and
$\sM_{\rm G}^{-1}\sV_{\rm G}^{1/3}=20/23, 20/22, 20/21$,
$20/20, 20/19, 20/18$, from bottom to top.

An inspection of Fig.\,\ref{f:hhTchi} discloses weak dependence of reduced
macrotemperature on  the ratio, $\Xi_{\rm G}/\Xi_{\rm P}$, within the range
considered.   Accordingly, reduced macrotemperature might be conceived as
independent of (infinite) scaled truncation radii to a first extent.

\subsection{Critical curves}
\label{crit}

\begin{figure*}[t]
\begin{center}
\includegraphics[scale=0.8]{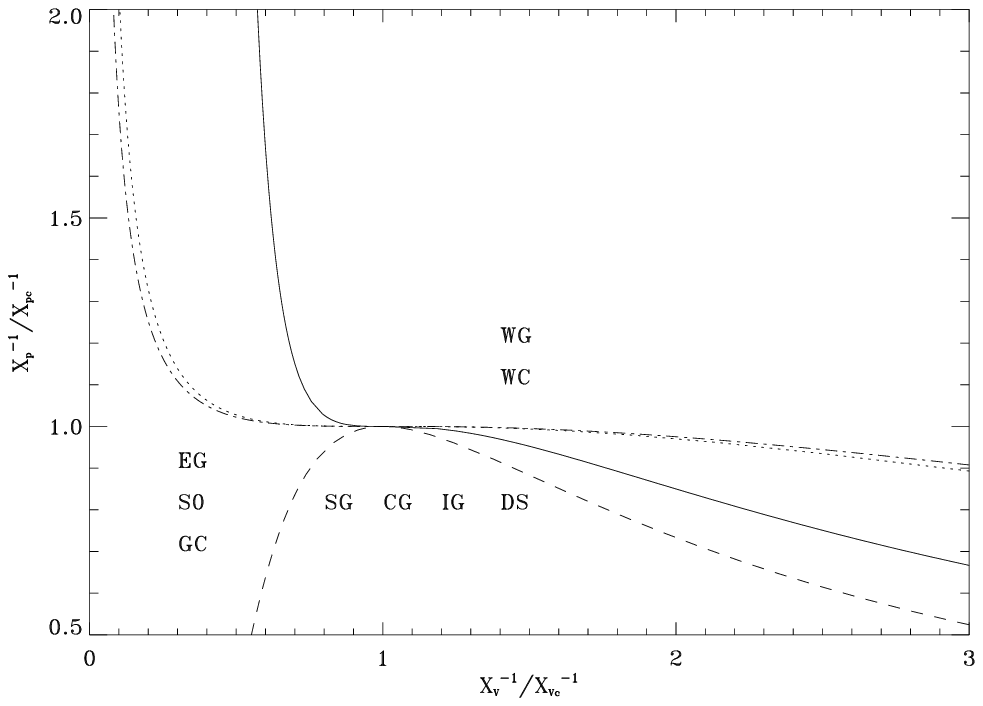}
\caption{Comparison between VDW critical isothermal (full), HH
critical fractional isoenergetic (dotted), and HN/NH critical fractional
isoenergetic (dot-dashed).   With regard to ordinary fluids, vapour and
liquid phase coexist within the bell-shaped region bounded by the dashed curve
and, in addition, $X_{\rm V}^{-1}=V, X_{\rm P}^{-1}=p$.   More extended
(along the horizontal direction) bell-shaped regions are expected for both
HH and HN/NH fractional isoenergetics.   The critical point, by definition,
reads $(\sX_{\rm V_c},\sX_{\rm p_c},\sX_{\rm T_c})\equiv(1,1,1)$.   Different
letters denote expected location of different
astrophysical systems.   Caption: EG - elliptical galaxies; S0 - lenticular
galaxies; SG - spiral galaxies including barred; IG - irregular galaxies; DS -
dwarf spheroidal galaxies; GC - globular clusters; CG - galaxy clusters; WC -
wholly gaseous clouds where stars never formed; WG - (hypothetical)
wholly gaseous galaxies where stars never formed.}
\label{f:mris101}
\end{center}
\end{figure*}

VDW critical isothermal and HH, HN/NH, critical fractional isoenergetic
are compared in Fig.\,\ref{f:mris101}, where the dashed curve is the same as
in Fig.\,\ref{f:vris100}.   Accordingly, vapour and liquid phase of ordinary
fluids coexist within the bell-shaped region bounded by the dashed curve.
Both HH and HN/NH critical fractional isoenergetic are more extended along
horizontal direction with respect to VDW critical isothermal, which implies a
more flattened counterpart of the above mentioned bell-shaped region.   The
critical point, by definition, reads
$(\sX_{\rm V_c},\sX_{\rm p_c},\sX_{\rm T_c})\equiv(1,1,1)$.

\section{Discussion}
\label{disc}

Tidal interactions between neighbourhing
bodies span across the whole admissible
range of lengths in nature: from, say,
atoms and molecules to galaxies and
clusters of galaxies i.e. from micro to
macro cosmos.   Ordinary fluids are collisional,
which makes the stress tensor be isotropic
and the velocity distribution obey
Maxwell's law.   Tidal interactions
(electromagnetic in nature) therein act
between colliding particles (e.g., LL67,
Chap.\,\,VII, \S74).   Astrophysical fluids
could be collisionless, which makes the stress
tensor be anisotropic and the velocity
distribution no longer obey Maxwell's
law.   Tidal interactions (gravitational
in nature) therein act between single
particles and the system as a whole (e.g.,
C10).

In both cases, an equation of state can
be formulated in reduced variables: the
VDW equation for ordinary fluids and an
equation which depends on density
profiles for astrophysical fluids.
For sufficiently mild density profiles,
fractional isoenergetics are characterized
by the occurrence of two extremum points,
similarly to isothermals where a
transition from liquid to gaseous phase
takes place, or vice versa.   For
sufficiently steep density profiles, the
critical fractional isoenergetic exhibits
a single horizontal inflexion point,
which defines the critical point.
Fractional isoenergetics below and above
the critical one, show two or no extremum
point, respectively, in complete analogy
with VDW isothermals.   In any
case, the existence of an equation of
state in reduced variables implies the
validity of the law of corresponding
states for macrogases with assigned
density profiles.

For astrophysical fluids, the existence
of a phase transition must necessarily
be assumed as a working hypothesis by
analogy with ordinary fluids.   The
phase transition has to be conceived
between gas and stars, and the (${\sf O}
\sX_V^{-1}\sX_p^{-1}$) plane may be divided into
three parts, namely
\begin{description}
\item[(i)]
a region bounded
by the critical fractional isoenergetic on
the left of the critical point, and 
the locus of onset of phase transition
on the right of the critical point, where
only gas exists;
\item[(ii)]
a region bounded by
the critical fractional isoenergetic on
the left of the critical point,
the locus of onset of phase transition
on the left of the critical point, and
the vertical axis, where only stars exist;
\item[(iii)]
a region bounded by
the locus of onset of phase transition,
and the horizontal axis, where gas and
stars coexist.
\end{description}
The locus of onset of phase transition,
not shown in Fig.\,\ref{f:mris101} for
reasons explained above, is similar to
its counterpart related to ordinary fluids,
represented by the bell-shaped curve in
Fig.\,\ref{f:mris101}, but more extended
along the horizontal direction.

In this view, elliptical and S0 galaxies
lie on (ii) region unless hosting hot
interstellar gas, and the same holds for
globular clusters; spiral, irregular,
and dwarf spheroidal galaxies lie on
(iii) region, and the same holds for
cluster of galaxies; gas clouds where stars never formed lie on
(i) region, and the same holds for
hypothetical galaxies where stars never formed.

\section{Conclusion}
\label{conc}

Van der Waals' two
great discoveries, more specifically
a gas equation of state where tidal
interactions between molecules are
taken into account and the law of
corresponding states, related to
microcosmos, find a counterpart
with regard to macrocosmos.
After more than a century since the awarding
of the Nobel Prize in Physics, van
der Waals' ideas are still valid and
helpful to day for a full understanding
of the universe.

%\section{Acknowledgements}
%The author is indebted to an anonymous referee for helpful
%comments which  improved an
%earlier version of the manuscript.
%Thanks are due to T. Valentinuzzi for fruitful discussions.
%The analytical integrations needed in the current paper
%were helped substantially by use of the Mathematica
%package and visiting the internet site:
%``HTTP://INTEGRALS. WOLFRAM.COM/INDEX.CGI''.   This
%is why we are deeply grateful to the Wolfram staff,
%in particular to Daniel Lichtblau, and wish to
%acknowledge all the facilities encountered
%therein.
%\newpage

\appendix
\section*{Appendix}

\section{Additional features of VDW isothermals \\
         on the Clapeyron reduced plane}
\label{a:expo}

The equation of a generic VDW isothermal
on the Clapeyron reduced plane, $({\sf O}\sV\sP)$,
is (e.g., LL67, Chap.\,\,VIII, \S 85; C10):
\begin{lefteqnarray}
\label{eq:rW2}
&& \sP=\frac{8\sT}{3\sV-1}-\frac3{\sV^2}~~;
\end{lefteqnarray}
and the first and the second derivative
with respect to $\sV$ read:
\begin{lefteqnarray}
\label{eq:rW3}
&& \left(\frac{\partial\sP}{\partial\sV}\right)_{\sT}=-\frac{24\sT}
{(3\sV-1)^2}+\frac6{\sV^3}~~; \\
\label{eq:rW4}
&& \left(\frac{\partial^2\sP}{\partial\sV^2}\right)_{\sT}=\frac
{144\sT}{(3\sV-1)^3}-\frac{18}{\sV^4}~~;
\end{lefteqnarray}
where, for assigned $\sT$, the domain of the function,
$\sP(\sV)$, is $\sV>1/3$, $\sV=1/3$ is a vertical
asymptote, and $\sP=0$ is a horizontal asymptote.
In the special case of the critical point, $(\sV,\sT,\sP)
\equiv(1,1,1)$, the above mentioned derivatives are null,
as expected.

The extremum points, via Eq.\,(\ref{eq:rW3}), are
defined by the relation:
\begin{equation}
\label{eq:ext}
f(\sV)=\frac{(3\sV-1)^2}{4\sV^3}=\sT~~;
\end{equation}
which is satisfied on the critical point, as
expected.   The function on the left-hand side
of Eq.\,(\ref{eq:ext}) has two extremum points:
a minimum at $\sV=1/3$ (outside the physical
domain) and a maximum at $\sV=1$, where $\sT=1$.
Accordingly, Eq.\,(\ref{eq:ext}) is never
satisfied for $\sT>1$, which implies no extremum
point for related isothermals, as expected.
The contrary holds for $\sT<1$, where it can be
seen that the third-degree equation associated
to Eq.\,(\ref{eq:rW3}) has three real solutions,
related to extremum points.   One lies outside
the physical domain, which implies $\sV\le1/3$.
The remaining two are obtained as the intersections
between the curve, $y=f(\sV)$, expressed by Eq.\,(\ref
{eq:ext}), and the straight line, $y=\sT$, keeping
in mind that $f(1/3)=0$, $f(1)=1$, and $\lim_{\sV\to
+\infty}f(\sV)=0$.

The third-degree equation associated to Eq.\,(\ref
{eq:rW3}), may be ordered as:
\begin{leftsubeqnarray}
\slabel{eq:3dea}
&& \sV^3-9a\sV^2+6a\sV-a=0~~; \\
\slabel{eq:3deb}
&& a=\frac1{4\sT}~~;
\label{seq:3de}
\end{leftsubeqnarray}
where, with regard to the standard formulation (e.g.,
Spiegel, 1968, Chap.\,9):
\begin{equation}
\label{eq:3dx}
x^3+a_1x^2+a_2x+a_3=0~~;
\end{equation}
the discriminants of Eq.\,(\ref{eq:3dea}) are:
\begin{lefteqnarray}
\label{eq:Q}
&& Q=\frac{3a_2-a_1^2}9=a(2-9a)~~; \\
\label{eq:R}
&& R=\frac{9a_1a_2-27a_3-2a_1^3}{54}=\frac{a(1-18a+54a^2)}2~~; \\
\label{eq:D}
&& D=Q^3+R^2=\frac{a^2(1-4a)}4~~;
\end{lefteqnarray}
where $D=0$ in the special case of the critical
isothermal $(\sT=1, a=1/4)$, $D<0$ for
$\sT<1$, and $D>0$ for $\sT>1$.   Accordingly,
three (at least two coincident) real solutions exist if $D=0$,
three different real solutions
if $D<0$, one real and two complex coniugate if $D>0$.
A real solution, $\sV_0$, always lies outside
the physical domain.

The three real solutions $(D\le0)$ may be expressed as
(e.g., Spiegel, 1968, Chap.\,9):
\begin{leftsubeqnarray}
\slabel{eq:rsola}
&& \sV_1=2\sqrt{-Q}\cos\left(\pi+\frac\theta3+\frac{0\pi}3\right)-
\frac13a_1~~; \\
\slabel{eq:rsolb}
&& \sV_2=2\sqrt{-Q}\cos\left(\pi+\frac\theta3+\frac{2\pi}3\right)-
\frac13a_1~~; \\
\slabel{eq:rsolc}
&& \sV_3=2\sqrt{-Q}\cos\left(\pi+\frac\theta3+\frac{4\pi}3\right)-
\frac13a_1~~; \\
\slabel{eq:rsold}
&& \theta=\arctan\left(\frac{\sqrt{-D}}R\right)~~;
\label{seq:rsol}
\end{leftsubeqnarray}
where $a_1=-9a$ and, in the special case of
critical isothermal, $a=1/4$, $Q=-1/16$, $R=-1/64$,
$D=0$, which implies $\sV_0=\min(\sV_1,\sV_2,\sV_3)$,
$\sV_{\rm A}=\sV_{\rm B}=\sV_{\rm C}=\sV_{\rm D}=\sV_{\rm E}=
\max(\sV_1,\sV_2,\sV_3)$.   A null factor appears in
Eq.\,(\ref{eq:rsola}) to save aesthetics.
In the special case,
$\sT\to0$, Eq.\,(\ref{eq:3dea}) reduces to a
second-degree equation whose solutions are
$\sV_{01}=\sV_{02}=1/3$, while the related
function is otherwise divergent as $a\to+\infty$.
In general, the extremum points of VDW isothermals
$(\sT\le1)$ occur at $\sV=\sV_{\rm B}$ (minimum) and
$\sV=\sV_{\rm D}$ (maximum), $\sV_{\rm B}\le\sV_{\rm D}$.   As
$\sT\to0$, $\sV_{\rm B}\to1/3$, $\sV_{\rm D}\to+\infty$,
where, in all cases, $1/3<\sV_{\rm B}\le1\le\sV_{\rm D}$.

\section{Intersections between real and VDW \\ isothermals of equal
temperature}
\label{a:inrv}

With regard to selected real and VDW isothermals of equal temperature,
%on the Clapeyron reduced plane, $({\sf O}\sV\sP)$,
three distinct
intersections occur in presence of saturated
vapour i.e. below the critical temperature,
which are coincident at the critical temperature.
Vapour pressure maintains
constant in presence of liquid phase i.e.
$\sV_{\rm A}\le\sV\le\sV_{\rm E}$, where an
infinitesimal vapour and liquid mass fraction
characterizes the extreme reduced volume values,
$\sV_{\rm A}$ and $\sV_{\rm E}$, respectively.
In presence of a sole phase, liquid $(\sV<\sV_{\rm A})$
or gas $(\sV>\sV_{\rm E})$, related
real and VDW
isothermal branches coincide.

The intersections between real and VDW
isothermals of equal temperature
(within the range where they are different)
can be determined as intersections between
horizontal lines and VDW isothermals
on the Clapeyron reduced plane.   Two
intersections necessarily occur at
$(\sV_{\rm A},\sP_{\rm A})$ and
$(\sV_{\rm E},\sP_{\rm E})$.   The third one,
$(\sV_{\rm C},\sP_{\rm C})$, necessarily lies
between the other two where, in addition,
$\sP_{\rm A}=\sP_{\rm C}=\sP_{\rm E}$.
The coordinates of the extremum points,
$(\sV_{\rm B},\sP_{\rm B})$, minimum, and
$(\sV_{\rm D},\sP_{\rm D})$, maximum,
%$\sV_{\rm B}\le\sV_{\rm D}$,
%$\sP_{\rm B}\le\sP_{\rm D}$,
must necessarily satisfy the following
inequalities: 
$\sV_{\rm A}\le\sV_{\rm B}\le\sV_{\rm C}$;
$\sV_{\rm C}\le\sV_{\rm D}\le\sV_{\rm E}$;
$\sP_{\rm B}\le\sP_{\rm C}\le\sP_{\rm D}$.
It can be seen that regions, ${\sf ABC}$,
${\sf CDE}$, bounded by real and VDW isothermals with equal temperature,
have equal areas (e.g., LL67, Chap.\,\,VIII, \S 85), which yields
the following relation (e.g., C10):
\begin{equation}
\label{eq:S12}
\sP_{\rm C}=\frac83\frac{\sT}{\sV_{\rm E}-\sV_{\rm A}}\ln\frac{3\sV_{\rm E}-1}{3\sV_{\rm A}-1}-\frac3
{\sV_{\rm A}\sV_{\rm E}}~~;
\end{equation}
where, for a selected isothermal, the unknowns
are $\sP_{\rm C}=\sP_{\rm A}=\sP_{\rm E}$, $\sV_{\rm A}$, and $\sV_{\rm E}$.

The reduced volumes, $\sV_{\rm A}$, $\sV_{\rm C}$, $\sV_{\rm E}$,
see Fig.\,\ref{f:vrar100}, may be considered as
intersections between a VDW isothermal
$(\sT<1)$ and a horizontal straight line, $\sP=\sP_{\rm C}$,
on the Clapeyron reduced plane.
% $({\sf O}\sV\sP)$ plane.
In other words,
$\sV_{\rm A}$, $\sV_{\rm C}$, $\sV_{\rm E}$, are the real solutions
of the third-degree equation:
\begin{equation}
\label{eq:3Wrr}
\sV^3-\left(\frac13+\frac83\frac{\sT}{\sP_{\rm C}}\right)\sV^2+\frac3
{\sP_{\rm C}}\sV-\frac1{\sP_{\rm C}}=0~~;
\end{equation}
which has been deduced from Eq.\,(\ref{eq:rW2}),
particularized to $\sP=\sP_{\rm C}$.   Related
solutions may be calculated using Eqs.\,(\ref{seq:rsol}).
The last unknown, $\sP_{\rm C}$, is determined from Eq.\,(\ref
{eq:S12}).

An inspection of Fig.\,\ref{f:vrar100} shows that
the points, {\sf A} and {\sf E}, are located
on the left of the minimum, {\sf B}, and on
the right of the maximum, {\sf D}, respectively.
Keeping in mind the above results, the following
inequality holds: $\sV_{\rm A}\le\sV_{\rm B}\le1\le\sV_{\rm D}\le
\sV_{\rm E}$, which implies further investigation on
the special case, $\sV_{\rm C}=1$.   The particularization
of VDW equation of state, Eq.\,(\ref{eq:rW2}), to the
point, ${\sf C}={\sf C_1}$, assuming $\sV_{\rm C_1}=1$,
yields:
\begin{equation}
\label{eq:TVC1}
\sT=\frac{\sP_{\rm C_1}+3}4~~;
\end{equation}
and Eq.\,(\ref{eq:3Wrr}) reduces to:
\begin{leftsubeqnarray}
\slabel{eq:3dba}
&& \sV^3-(1+2b)\sV^2+3b\sV-b=0~~; \\
\slabel{eq:3dbb}
&& b=\frac1{\sP_{\rm C_1}}~~;
\label{seq:3db}
\end{leftsubeqnarray}
with regard to the generic third-degree equation,
Eq.\,(\ref{eq:3dx}), the three solutions, $x_1$,
$x_2$, $x_3$, satisfy the relations (e.g., Spiegel,
1968, Chap.\,9):
\begin{leftsubeqnarray}
\slabel{eq:x123a}
&& x_1+x_2+x_3=-a_1~~; \\
\slabel{eq:x123b}
&& x_1x_2+x_2x_3+x_3x_1=a_2~~; \\
\slabel{eq:x123c}
&& x_1x_2x_3=-a_3~~;
\label{seq:x123}
\end{leftsubeqnarray}
where, in the case under discussion:
\begin{leftsubeqnarray}
\slabel{eq:b123a}
&& a_1=-1-2b~~;\qquad a_2=3b~~;\qquad a_3=-b~~; \\
\slabel{eq:b123b}
&& x_1=\sV_{\rm A}~~;\qquad x_2=\sV_{\rm C_1}=1~~;\qquad x_3=\sV_{\rm E}~~;
\label{seq:b123}
\end{leftsubeqnarray}
and the substitution of Eqs.\,(\ref{seq:b123})
into two among (\ref{seq:x123}) yields:
\begin{leftsubeqnarray}
\slabel{eq:VAEa}
&& \sV_{\rm A}=b-\sqrt{b^2-b}~~; \\
\slabel{eq:VAEb}
&& \sV_{\rm E}=b+\sqrt{b^2-b}~~;
\label{seq:VAE}
\end{leftsubeqnarray}
finally, the combination of Eqs.\,(\ref{eq:TVC1}),
(\ref{eq:3dbb}), and (\ref{seq:VAE}) produces:
\begin{leftsubeqnarray}
\slabel{eq:VAETa}
&& \sV_{\rm A}=\frac{1-2\sqrt{1-\sT}}{4\sT-3}~~;\qquad\sT\le1~~; \\
\slabel{eq:VAETb}
&& \sV_{\rm E}=\frac{1+2\sqrt{1-\sT}}{4\sT-3}~~;\qquad\sT\le1~~;~~;
\label{seq:VAET}
\end{leftsubeqnarray}
which, together with $\sV_{\rm C_1}=1$, are the abscissae
of the intersection points between a selected VDW
isothermal on the Clapeyron reduced
%$({\sf O}\sV\sP)$
plane and
the straight line, $\sP=\sP_{\rm C_1}$, in the special case
under discussion.

The substitution of Eqs.\,(\ref{seq:VAET}) into
(\ref{eq:S12}), the last related to $\sP=\sP_{\rm C_1}$ via
Eq.\,(\ref{eq:TVC1}), yields:
\begin{equation}
\label{eq:S12C}
\frac{\sT}{\sqrt{1-\sT}}\ln\frac{3-2\sT+3\sqrt{1-\sT}}{3-2\sT-3\sqrt{1-\sT}}
=6~~;
\end{equation}
which, keeping in mind the limit:
\begin{equation}
\label{eq:Slim}
\lim_{x\to0}\left[\frac1x\ln\frac{1+x}{1-x}\right]=2~~;
\end{equation}
holds only for the critical isothermal,
$\sT=1$.   Accordingly, the abscissa of the intersection
point, {\sf C}, between a selected VDW isothermal
and related real isothermal, see Fig.\,\ref{f:vrar100},
cannot occur at $\sV_{\rm C}=1$ unless the critical isothermal is considered.
Then the third-degree equation,
Eq.\,(\ref{eq:3Wrr}), must be solved in the general case $(\sV_{\rm C}\ne1)$
by use of  Eqs.\,(\ref{seq:rsol}).

\section{The limit of zero absolute temperature}
\label{a:zero}

Ideal gas equation of state, expressed by Eq.\,(\ref{eq:gid}),
%as:
%\begin{displaymath}
%pV=NkT~~;
%\end{displaymath}
in the limit of zero absolute temperature, $T=0\,$K, implies (i) pressure
attains any value, $p>0$, provided $V\to0$, and (ii) volume attains any value,
$V>0$, provided $p\to0$.   Accordingly, related ideal isothermal on the
Clapeyron plane, $({\sf O}Vp)$, tends to positive coordinate semiaxes.   The
same holds for reduced variables on the Clapeyron reduced plane,
$({\sf O}\sV\sP)$.

VDW equation of state, expressed by Eq.\,(\ref{eq:VdW}),
%as:
%\begin{displaymath}
%\left(pA\frac{N^2}{V^2}\right)(V-NB)=NkT~~;
%\end{displaymath}
in the limit of zero absolute temperature, $T=0\,$K, implies (i) pressure
attains any value, $p>-A/B^2$, provided $V\to NB$, and (ii) volume attains
any value, $V>NB$, provided $p\to-A/B^2$; where $NB=V_{\rm c}/3$;
$A/B^2=27p_{\rm c}$; via Eqs.\,(\ref{eq:Vc}); (\ref{eq:pc}); respectively.
Accordingly, related VDW isothermal on the Clapeyron plane,
$({\sf O}Vp)$, tends to subdomains, $V>V_{\rm c}/3$; $p>-27p_{\rm c}$; or, in
reduced variables, $\sV>1/3$; $\sP>-27$; respectively.

With regard to VDW isothermal in the limit of absolute zero
reduced temperature, $\sT\to0$, extremum points occur at $\sV_{\rm B}\to1/3$
(minimum) and $\sV_{\rm D}\to+\infty$ (maximum), as inferred in Appendix
\ref{a:expo} and, in addition, $\sP_{\rm B}\to-27$ (minimum) and
$\sP_{\rm D}\to0$ (maximum), according to above considerations.   In summary,
${\sf B}\to{\sf B_0}\equiv(1/3,-27)$ and
${\sf D}\to{\sf D_0}\equiv(+\infty,0)$.

The real isothermal $(\sV_{\rm A}\le\sV\le\sV_{\rm E})$ in the limit
of absolute zero reduced temperature, $\sT\to0$, can be determined via
Eq.\,(\ref{eq:S12})
%as:
%\begin{displaymath}
%\sP_{\rm C}=\frac83\frac{\sT}{\sV_{\rm E}-\sV_{\rm A}}\ln\frac{3\sV_{\rm E}-1}{3\sV_{\rm A}-1}-\frac3
%{\sV_{\rm A}\sV_{\rm E}}~~;
%\end{displaymath}
keeping in mind $\sV_{\rm B}\ge\sV_{\rm A}$ and $\sV_{\rm D}\le\sV_{\rm E}$.
The result is $\sP_{\rm C}\to0$.

In general, intersections between VDW isothermals and horizontal
lines, $\sP=\sP_{\rm C}$, on the  Clapeyron reduced plane, $({\sf O}\sV\sP)$,
are solution of a third-degree equation expressed by Eq.\,(\ref{eq:3Wrr})
%as:
%\begin{displaymath}
%\sV^3-\left(\frac13+\frac83\frac{\sT}{\sP_{\rm C}}\right)\sV^2+\frac3
%{\sP_{\rm C}}\sV-\frac1{\sP_{\rm C}}=0~~;
%\end{displaymath}
which, in the special case of the horizontal axis, $\sP_{\rm C}=0$, reduces to
a second-degree equation, as:
\begin{equation}
\label{eq:2Wrr}
\frac83\sT\sV^2+3\sV-1=0~~;
\end{equation}
and related solutions read:
\begin{equation}
\label{eq:2srr}
\sV=\frac{9\mp\sqrt{81-96\sT}}{16\sT}~~;
\end{equation}
where real solutions imply $\sT\le81/96=27/32$ and the special case,
$\sT=27/32$, yields a VDW isothermal which is tangent to the horizontal axis
on $(\sV_{\rm B},0)$.    If, in particular, $\sP_{\rm C}=0$ and
$\sT\to0$, Eq.\,(\ref{eq:3Wrr}) reduces to a first-degree equation, as:
\begin{equation}
\label{eq:1Wrr}
3\sV-1=0~~;
\end{equation}
and related solution reads $\sV=1/3$, as expected.

To get further insight, let a variable, $x$, be defined as:
\begin{equation}
\label{eq:x2}
81-96\sT=x^2~~;\qquad\sT=\frac{81-x^2}{96}~~;\qquad0\le x<9~~;
\end{equation}
where the limit of zero absolute reduced temperature, $\sT\to0^+$, relates to
$x\to9^-$.
The substitution of Eq.\,(\ref{eq:x2}) into (\ref{eq:2srr}) yields:
\begin{equation}
\label{eq:2sr0}
\sV=\frac{9\mp x}{16(81-x^2)/96}=\frac{96}{16}\frac{9\mp x}{(9+x)(9-x)}=\frac6
{9\pm x}~~;
\end{equation}
with regard to VDW isothermals, $\sT=(81-x^2)/96$.  In the limit
of zero absolute reduced temperature, $x\to9^-$, the above solutions read
$\sV\to6/18=1/3$ and $\sV\to+\infty$, as expected.

\begin{figure*}[t]
\begin{center}
\includegraphics[scale=0.8]{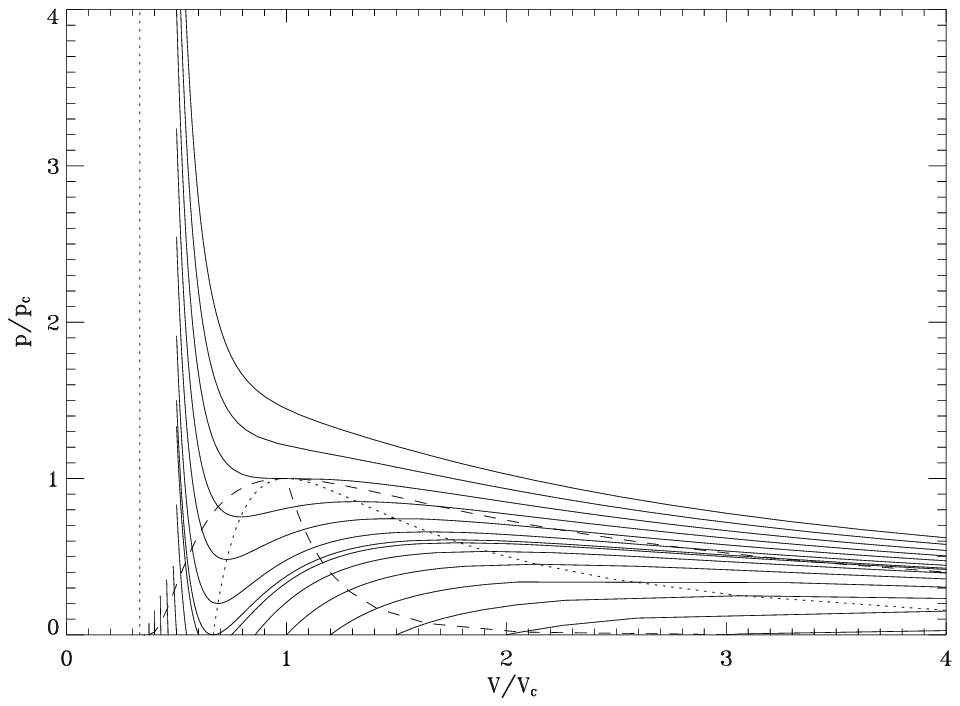}
\caption{VDW isothermals, $\sT=(81-x^2)/96$, plotted for integer
$x$, $0\le x\le8$, from top to bottom where the first is tangent to the
horizontal axis and the last lies outside the box.   The region of negative
pressure or sufficiently small volume is not shown for
sake of clarity.   In the limit of zero absolute reduced temperature,
$\sT\to0^+$, $x\to9^-$, VDW isothermal reads $\sP>-27;$
$\sV\to1/3;$ and $\sP\to-27;$ $\sV>1/3$. Other curves and captions as in
Fig.\,\ref{f:vris100}.}
\label{f:vris101}
\end{center}
\end{figure*}

VDW isothermals are plotted in Fig.\,\ref{f:vris101} for integer
$x$, $0\le x\le8$, where values related to negative pressure or sufficiently
small volume are not shown for sake of clarity.
An inspection of Fig.\,\ref{f:vris101} shows regions, ${\sf ABC}$,
${\sf CDE}$, bounded by real and VDW reduced isothermals
(e.g., Fig.\,\ref{f:vrar100}), exhibit equal areas (e.g., LL67, Chap.\,VII,
\S85) which increase as reduced temperature decreases and tend to infinite as
$\sT\to0$.

More specifically, the trend is described by the following relations:
\begin{eqnarray*}
&& \sV_{\rm A}\to\frac13~;~\sV_{\rm B}\to\frac13~;~\sV_{\rm C}\to+\infty~;~
\sV_{\rm D}\to+\infty~;~\sV_{\rm E}\to+\infty~;~\sT\to0~;~~ \\
&& \sP_{\rm A}\to0~;~~\sP_{\rm B}\to-27~;~~\sP_{\rm C}\to0~;~~
\sP_{\rm D}\to0~;~~\sP_{\rm E}\to0~;~~\sT\to0~;~~ 
\end{eqnarray*}
according to above considerations where, in general (e.g.,
Fig.\,\ref{f:vrar100}):
\begin{eqnarray*}
&& \sV_{\rm A}\le\sV_{\rm B}\le\sV_{\rm C}\le\sV_{\rm D}\le\sV_{\rm E}~~;\quad
\sT\le1~~;~~ \\
&& \sP_{\rm B}\le\sP_{\rm A}=\sP_{\rm C}=\sP_{\rm E}\le\sP_{\rm D}~;~~
\sT\le1~;~~
\end{eqnarray*}
according to the results of Appendix \ref{a:inrv}.

The limit of zero absolute reduced temperature, $\sT\to0$, implies (i) region
${\sf ABC}$ tends to a rectangular triangle of catheti,
$\overline{\sf AB}=\sP_{\rm A}-\sP_{\rm B}=27$ and
$\overline{\sf AC}=\sV_{\rm C}-\sV_{\rm A}\to+\infty$; (ii) region ${\sf CDE}$
tends to a triangle of height,
$\sP_{\rm D}-\sP_{\rm C}=\sP_{\rm D}-\sP_{\rm E}\to0$, and basis,
$\sV_{\rm E}-\sV_{\rm C}\to+\infty$; and (iii) regions ${\sf ABC}$ and
${\sf CDE}$ exhibit infinite area of same order, where their ratio equals
unity.

\section{Fractional macroisothermals}
\label{a:remi}

In considering whether fractional macroisothermals, $Y_{\rm T}=(T_{\rm P}/
T_{\rm G})=(\overline m_{\rm P}/\overline m_{\rm G})$ $(\phi/m)={\rm const}$
via
Eqs.\,(\ref{eq:YpVT}) and (\ref{eq:TPG}), hence $\phi/m={\rm const}$, are a
viable alternative with respect to fractional isoenergetics,
$X_{\rm T}=\phi={\rm const}$ via Eqs.\,(\ref{eq:msp}) and (\ref{eq:XpVT}), for
determining isoenergetics or macroisothermals,
attention shall be restricted in finding a counterexample.

With regard to UU macrogases, the following relation holds (C10):
\begin{equation}
\label{eq:phim}
\frac\phi m=\cases{
\frac{(2m+5)y^2-3}{2(y^3+m)}~~;   & $1\le y<+\infty~~;$      \cr
\frac{2(y^3+m)}{(2+5m)y-3my^3}~~; & $0<y\le1~~;$             \cr
}
\end{equation}
where $y=1$ implies $\phi/m=1$, independent of $m$, according to
Eq.\,(\ref{eq:mfy1}).  Related fractional macroisothermal reads:
\begin{leftsubeqnarray}
\slabel{eq:phm1a}
&& (2m+5)y^2-3=2(y^3+m)~~;\qquad1\le y<+\infty~~; \\
\slabel{eq:phm1b}
&& 2(y^3+m)=(2+5m)y-3my^3~~;\qquad0<y\le1~~;
\label{seq:phm1}
\end{leftsubeqnarray}
which, after some algebra, may be cast under the form:
\begin{equation}
\label{eq:mphi}
m=\cases{
\frac{2y^3-5y^2+3}{2(y^2-1)}=\frac{2y^2-3y-3}{2(y+1)}~~; & $1\le y<+\infty~~;$
                                                                           \cr
\frac{2y(1-y^2)}{3y^3-5y+2}=\frac{-2y(y+1)}{3y^2+3y-2}~~; & $0<y\le1~~;$   \cr
}
\end{equation}
and the function, $m=m(y)$, can be studied.

Special values are the following:
\begin{equation}
\label{eq:mphs}
m(0)=0~~;\qquad m(1)=-1~~;\qquad\lim_{y\to+\infty}m(y)=\lim_{y\to+\infty}y=+
\infty~~;
\end{equation}
where, in addition to the origin, zeroes of $m(y)$ are solutions of the
second-degree equation, $2y^2-3y-3=0$, hence $y=(3\mp\sqrt{9+24})/4$, and the
zero within the domain, $y\ge1$, reads:
\begin{equation}
\label{eq:mph0}
m\left(\frac{3+\sqrt{33}}4\right)=0~~.
\end{equation}

On the other hand, vertical asymptotes relate to solutions of the
second-degree equation, $3y^2+3y-2=0$, hence $y=(-3\mp\sqrt{9+24})/6$, and the
vertical asymptote within the domain, $0\le y\le1$, reads:
\begin{equation}
\label{eq:mphv}
\lim_{y\to y_0^\mp}m(y)=\pm\infty~~;\qquad y_0=\frac{-3+\sqrt{33}}6~~.
\end{equation}

The above results define the sign of $m(y)$ all over the domain, as:
\begin{leftsubeqnarray}
\slabel{eq:mpsa}
&& 0\le y\le\frac{\sqrt{33}-3}6~~;\qquad m(y)\ge0~~; \\
\slabel{eq:mpsb}
&& \frac{\sqrt{33}-3}6\le y\le\frac{\sqrt{33}+3}4~~;\qquad m(y)\le0~~; \\
\slabel{eq:mpsc}
&& \frac{\sqrt{33}+3}4\le y<+\infty~~;\qquad m(y)\ge0~~;
\label{seq:mps}
\end{leftsubeqnarray}
accordingly, the mass ratio, $m$, can assume both signs.

But $m\ge0$ by definition, which implies homodirection axis
ratios, $y$, within the range defined by Eq.\,(\ref{eq:mpsb}), cannot occur
along the fractional macro\-isothermal, $\phi/m=1$.   For this reason, it
would
be better dealing with fractional isoenergetics, where the above mentioned
inconvenient does not take place.

\end{document}